\documentclass[12pt,english]{article}
\pdfoutput=1
\usepackage{lmodern}
\usepackage[T2A,T1]{fontenc}
\usepackage[utf8]{inputenc}
\usepackage{geometry}
\usepackage{color}
\usepackage{babel}
\usepackage{mathtools}
\usepackage{amsmath}
\usepackage{amssymb}
\usepackage{graphicx}
\usepackage{subcaption}
\usepackage{textcomp}
\usepackage[numbers,sort&compress]{natbib}
\usepackage[unicode=true,pdfusetitle,
bookmarks=true,bookmarksnumbered=false,bookmarksopen=false,
breaklinks=false,pdfborder={0 0 1},backref=false,colorlinks=true]
{hyperref}
\hypersetup{
	citecolor=black,linkcolor=black,urlcolor=black}
\makeatletter

\@ifundefined{textcolor}{}
{
	\definecolor{BLACK}{gray}{0}
	\definecolor{WHITE}{gray}{1}
	\definecolor{RED}{rgb}{1,0,0}
	\definecolor{GREEN}{rgb}{0,1,0}
	\definecolor{BLUE}{rgb}{0,0,1}
	\definecolor{CYAN}{cmyk}{1,0,0,0}
	\definecolor{MAGENTA}{cmyk}{0,1,0,0}
	\definecolor{YELLOW}{cmyk}{0,0,1,0}
}
\usepackage{babel}
\makeatletter
\def\simgt{\mathrel{\lower2.5pt\vbox{\lineskip=0pt\baselineskip=0pt
			\hbox{$>$}\hbox{$\sim$}}}}
\def\simlt{\mathrel{\lower2.5pt\vbox{\lineskip=0pt\baselineskip=0pt
			\hbox{$<$}\hbox{$\sim$}}}}
\makeatother

\newcommand{\be}{\begin{equation}}
	\newcommand{\ee}{\end{equation}}

\newcommand{\tc}{\tilde{c}}
\newcommand{\tC}{\tilde{C}}

\newcommand{\inn}[2]{{#1}\cdot{#2}}

\newcommand{\mPl}{m_{\rm Pl}}

\newcommand{\BS}{\textrm{BS}}
\newcommand{\YM}{\textrm{YM}}
\newcommand{\GR}{\textrm{GR}}
\newcommand{\JBS}{\mathcal{J}^{a\tilde{a}}}
\newcommand{\JYM}{\mathcal{J}^{\mu a}}

\newcommand{\TGR}{\mathcal{T}^{\mu \nu}}
\newcommand{\Jdilaton}{\mathcal{J}}

\usepackage{array}   
\newcolumntype{L}{>{$}l<{$}}

\begin{document}
	\interfootnotelinepenalty=10000
	\baselineskip=18pt
	\hfill
	
	\vspace{2cm}
	\thispagestyle{empty}
	\begin{center}
		{\LARGE \bf 
			Gravitational Radiation \\
			\vspace{0.5cm}
			from Color-Kinematics Duality
		}
		\\
		\bigskip\vspace{1.cm}{
			{\large Chia-Hsien Shen}
		} \\[7mm]
		{\it  
			Mani L. Bhaumik Institute for Theoretical Physics, \\[-1mm]
			Department of Physics and Astronomy, UCLA, Los Angeles, CA 90095
		}\let\thefootnote\relax\footnote{e-mail:\url{chshen@physics.ucla.edu}

			} \\
	\end{center}
	\bigskip
	\centerline{\large \bf  Abstract}	
	
	\begin{quote} \small
		We perturbatively calculate classical radiation in Yang-Mills theory and dilaton gravity, to next-to-leading order in couplings.
		The radiation is sourced by the scattering of two relativistic massive scalar sources with the dynamical effect taken into account, corresponding to the post-Minkowskian regime in gravity.
		We show how to arrange the Yang-Mills radiation such that the duality between color and kinematics is manifest, including the three-term Jacobi identity.
		The search for duality-satisfying expressions exploits an auxiliary bi-adjoint scalar theory as a guide for locality.
		The double copy is obtained by replacing the color factors in Yang-Mills with kinematic counterparts, following Bern-Carrasco-Johansson construction in S-matrix.
		On the gravity side, the radiation is directly computed at the third post-Minkowskian order with massive sources.
		We find perfect agreement between observables in dilaton gravity and the Yang-Mills double copy.
		This non-trivially generalizes the leading-order rules by Goldberger and Ridgway.
		For the first time, the kinematic Jacobi identity appears beyond field-theory S-matrix, 
		suggesting that the color-kinematics duality holds more generally.
		Our results offer a path for simplifying analytical calculations in post-Minkowskian regime.
	\end{quote}

	\setcounter{footnote}{0}

	\newpage
	\setcounter{tocdepth}{2}
	\tableofcontents
	
	\newpage

\section{Introduction}
\label{sec:intro}
After a century of effort, gravity remains at the heart of modern physics.
The enormous complexity in gravity, even in the classical level, is one of the obstructions for making advances.
This can be seen from the non-linear nature in Einstein-Hilbert action which limits the space of exact solutions and
the range of perturbative orders. The need for a better theoretical understanding is particularly urgent, after the breakthrough of gravitational wave detection~\cite{Abbott:2016blz}.

Seeking insights for unraveling gravity from field theory has a long history. A classic example by Kawai, Lewellen and Tye (KLT)~\cite{KLT} is that a closed string can be related to a product of open strings. In the low-energy limit, the KLT relation implies that gravity can be viewed as a square of Yang-Mills theory (YM). Crucially, the hidden connection between gravity and YM is manifest at the level of the S-matrix, which is physical and free from the gauge artifacts.

The connection between gravity and YM has gone deeper after the seminal work by Bern, Carrasco, and Johansson (BCJ)~\cite{BCJ,Bern:2010ue}. They start with a YM amplitude whose
color factors $C_i$ of cubic diagrams are related by
\begin{gather}
C_i \pm C_j =0 \nonumber \\
C_i\pm C_j \pm C_k = 0
\label{eq:color_id_qft}
\end{gather}
from the anti-symmetry of structure constants and Jacobi identity.
A sharp conjecture by BCJ is that the kinematic numerators, when expanded in cubic diagrams, can be made to satisfy the same generic algebraic relations,
\begin{gather}
	C_i \pm C_j =0 \quad  \leftrightarrow \quad N_i \pm N_j =0 \nonumber \\
	C_i \pm C_j \pm C_k = 0 \quad \leftrightarrow \quad N_i \pm N_j \pm N_k = 0
\label{eq:kin_id_qft}
\end{gather}
which goes by the name \emph{color-kinematics duality}. 
Given their similarity, we replace color numerators $C_i$ with duality-satisfying representations $N_i$.
The resulting amplitude, with two copies of $N_i$, magically becomes a gravitational amplitude! 
By manifesting the duality between color and kinematics, the BCJ double copy
generalizes the KLT relations to quantum level.
Many gauge theories are shown to have color-kinematics duality, including various matter contents and supersymmetries at both tree and loop levels.
Their double copies agree with a variety of gravitational theories.
See Ref.~\cite{Carrasco:2015iwa,Cheung:2017pzi} and references therein.
Apart from conceptually intriguing, 
the complexity of gravity is tremendously simplified into YM degrees of freedom via the double copy, which is the key behind multi-loop gravity calculations~\cite{Bern:2009kd,Bern:2010ue,Bern:2010yg,Carrasco:2011mn,Bern:2012uf,Bern:2012cd,Bern:2012gh,Bern:2013yya,Bern:2013qca,Bern:2013uka,Johansson:2014zca,Bern:2014sna,Bern:2017yxu,Johansson:2017bfl,Bern:2017ucb,Bern:2018jmv}.

Can the double-copy idea really simplify calculation in gravity beyond the vanilla S-matrix, and possibly illuminate gravitational wave physics?
This is initiated in Ref.~\cite{Monteiro:2014cda} by studying the Kerr-Schild solutions in gravity which can be linked to gauge theory.
Such similarities between gravity and gauge theory have been extended further~\cite{Luna:2015paa,Ridgway:2015fdl,Luna:2016due,Cardoso:2016amd,Luna:2016hge,Bahjat-Abbas:2017htu,Carrillo-Gonzalez:2017iyj}.
The double-copy structure of classical radiation as Bremsstrahlung has been studied in various theories~\cite{Luna:2016due,Luna:2016hge,Goldberger:2016iau,Goldberger:2017frp,Luna:2017dtq,Goldberger:2017vcg,Goldberger:2017ogt,Chester:2017vcz,Li:2018qap}.
A squaring relation in curved spacetime among three-point amplitudes has been also found recently~\cite{Adamo:2017nia}.
All the studies so far have restricted to either known exact solutions, which are difficult to find in gravity, or to leading order in perturbative solutions. In contrast, the power of double copy in S-matrix goes far beyond the simplest examples at three point.
To show double copy is indeed promising, we have to establish an example beyond the leading-order approximation.

The classical radiation as perturbative solutions in worldline formalism is particularly promising for further studies \cite{Goldberger:2016iau,Goldberger:2017frp,Luna:2017dtq,Goldberger:2017vcg,Goldberger:2017ogt,Chester:2017vcz,Li:2018qap}. First, there is still a notion of on-shell emission amplitudes as gauge-invariant observables. Based on the success of S-matrix program, where the hidden connections are discovered from physical quantities, any double-copy structure should be revealed first in observables.
Second, the perturbative regime is standard in worldline formalism. The observables are available by brute-force calculation, so we can test the double-copy structure at higher orders.
Moreover, the worldline formalism directly gives results in the classical limit. Any progress made here can be readily applied to general relativity.

Taking classical limit is not as trivial as we might think. The meaning of classical limit is two-fold.
The obvious one is to set $\hbar \rightarrow 0$. A common belief is that classical limit corresponds to tree-level approximation.
While the statement is true for massless theories, it becomes subtle for massive particles because loop diagrams still contain classical pieces by the inhomogeneous scaling of propagators under the $\hbar \rightarrow 0$ limit~\cite{Iwasaki:1971vb,Holstein:2004dn}.
On top of this, the classical limit also forbids anti-particles from being present.
Given these two subtleties, it is non-trivial to extract classical limit from a field-theory calculation.
The endeavor of applying S-matrix techniques to general relativity is still an active area of research~\cite{BjerrumBohr:2002kt,Holstein:2008sx,Neill:2013wsa,Bjerrum-Bohr:2013bxa,Bjerrum-Bohr:2014lea,Bjerrum-Bohr:2014zsa,Bjerrum-Bohr:2016hpa,Bjerrum-Bohr:2017dxw,Cachazo:2017jef,Guevara:2017csg,Laddha:2018rle,Laddha:2018myi,Laddha:2018vbn,Bjerrum-Bohr:2018xdl}.
The worldline formalism is therefore advantageous because both can be implemented straightforwardly.

In this paper we calculate classical radiation in YM and dilaton gravity to next-to-leading order, with a particular focus on their double-copy structure.
The gravitational theory considered here is dilaton gravity, which contains a dilaton minimally coupled in gravity. It is the expected gravitational theory by double copying YM. 
The system is comprised of two massive scalar particles, with arbitrary mass ratio and initial velocities, as long as the impact parameter is much larger than Schwarzschild radius. (Details will be discussed in Section~\ref{sec:setups}.)
After the two point sources pass through each other, classical radiation is measured by an observer at spatial infinity.
This corresponds to the post-Minkowskian regime in gravity~\cite{Bertotti:1956,BERTOTTI1960169,Rosenblum:1978zr,Westpfahl:1979gu,Portilla:1979xx,Portilla:1980uz,Bel:1981be,Damour:1981bh,Westpfahl:1985,Westpfahl:1987},
where a scattering process is calculated order-by-order in Newton constant $G_N$.
The scattering angle at the second order has been computed by Westpfahl~\cite{Westpfahl:1985}.
From the field theory point of view, this is an inelastic scattering of two scalars with a graviton in the final state.
Our results should be comparable to the one-loop integrand in the aforementioned classical limit.
It should be distinguished from the post-Newtonian approximation of a bounded, rather than scattering, binary system where the perturbation is done around small velocities. (A suitable field-theory framework for post-Newtonian system would be the effective field theory constructed in Ref.~\cite{Goldberger:2004jt}.)
Although the scattering scenario might not occur as often as inspiral black holes in astrophysics, it has been applied as inputs to the effective one-body formalism which could be useful for simulating binary black holes~\cite{Damour:2016gwp,Bini:2017xzy,Vines:2017hyw,Damour:2017zjx}.

At next-to-leading order, the three-term Jacobi identity begins to appear among the color factors in YM.
Whether a kinematic version of Jacobi identity exists becomes a non-trivial test of color-kinematics duality.
Although the systems we investigated are still scattering processes, classical radiation has a very different structure than field-theory S-matrix, rooted from the non-trivial classical limit discussed above. 
In particular, the locality structure is scrambled under the classical limit. This makes double copy challenging because one of its pillars is the preservation of locality by maintaining the propagators.
To expose color-kinematics duality, we exploit an auxiliary bi-adjoint scalar theory (BS) as a guide.
The double copy construction is then formulated in a suitable fashion for worldline formalism.
Despite many difference in details, the kinematic numerators in YM can still be arranged to satisfy the same algebraic relation as color factors as Eq.~\eqref{eq:kin_id_qft}, including
\begin{equation}
	C_i\pm C_j \pm C_k = 0 \quad \leftrightarrow \quad N_i \pm N_j \pm N_k = 0
\end{equation}
as the first example of kinematic Jacobi identity beyond field-theory S-matrix. 
Despite all such complications, it is still striking that the core of color-kinematics duality in Eq.~\eqref{eq:kin_id_qft} remains to hold.

After establishing color-kinematics duality, we find the on-shell classical radiation agrees perfectly between dilaton gravity and the double copy of YM.
Already at the level of action, the complexity in gravity is much more significant than YM and BS.
Direct computation in gravity then leads to an explosion in the number of terms in the final expression.
However, the success of color-kinematics duality simplifies a complicated expression in the gravity amplitude $A_{\GR}$ as
\begin{equation}
{\large
	A_{\BS} \xrightarrow{\quad \tC_i \rightarrow N_i \quad}\quad A_{\YM} \quad \xrightarrow{\quad C_i \rightarrow N_i \quad} A_{\GR}}
\end{equation}
by using the bi-adjoint scalar amplitude $A_{\BS}$ supplemented with duality-satisfying representation $N_i$ from YM. 
This opens the avenue to extend the power of color-kinematics duality to perturbative solutions in classical gravity.

The rest of the paper is organized as follows. We introduce the backgrounds in Section~\ref{sec:setups}, including a review on worldline formalism and double copy construction. A summary of calculation is given in Section~\ref{sec:calculation}. Most of the next-to-leading order results are lengthy and thus are provided as ancillary files to this work's arXiv submission, \texttt{ym\_pos\_NLO.m}, \texttt{ym\_radiation\_NLO.m}, \texttt{gr\_pos\_NLO.m}, \texttt{gr\_dilaton\_NLO.m} and \texttt{gr\_radiation\_NLO.m}, whose definitions can be found in Appendix~\ref{app:def} and~\ref{app:worldline}.
After the calculation is done, we discuss color-kinematics duality in Section~\ref{sec:bcj}, including the BCJ and KLT amplitude relations.
We conclude in Section~\ref{sec:discussion}.

\section{Setups}
\label{sec:setups}
This section introduces the necessary backgrounds before we go to next-leading-order calculations. 
The first two sections contains a review of the worldline formalism used in Ref.~\cite{Goldberger:2016iau,Goldberger:2017frp}, in a fashion that is readily applicable for higher order calculations.
The last part is devoted to be a compact review of color-kinematics duality and double copy.

\subsection{Actions and Equations of Motion}
The systems we consider in this paper consist of point particles and radiation fields, which are described by worldlines and field theories respectively. For example, in Yang-Mills theory we have point particles with trajectory and color charge degrees of freedom along the worldlines, as well as the gauge field mediating the interaction. The dynamics is compactly encoded in the full action
\begin{equation}
S = S_{pp} + S_{\BS / \YM / \GR}
\nonumber
\end{equation}
as a combination of the point-particle action $S_{pp}$ and the field-theory part $S_{\BS / \YM / \GR}$, whose details depend on the theories.
While the worldline degrees of freedom only appear in $S_{pp}$, the fields couple to both the worldline and the fields themselves. 
The actions are usually fixed by power-counting and symmetries in the system.
Once the action is given, the classical equations of motion (EoM) follow straightforwardly.
Let us discuss YM, dilaton gravity, and finally BS in turn.

\paragraph{Yang-Mills Theory}
The full action in YM is the sum of
\begin{equation}
\begin{split}
S_{pp} & = -m\int d\tau +\int d\tau\, \psi^{\dagger} iv\cdot D \psi \\
S_{\textrm{YM}} & = -\frac{1}{4}\,\int d^dx\, F^{a\mu\nu} F^{a}_{\mu\nu},
\end{split}
\end{equation}
where $\tau$ is the proper time of the point particle and the proper velocity $v^\mu = dx^\mu/d\tau$. 
With the gauge field $A^a_\mu$ and coupling $g$, we also have the covariant derivative $D_\mu = \partial_{\mu} +ig A^a_\mu T^a$ and $F^a_{\mu\nu} = \partial_{[\mu} A^a_{\nu]} -g f^{abc} A^b_\mu A^c_\nu$ as the non-Abelian field strength.\footnote{Other relevant conventions are $[T^a,T^b]= if^{abc}T^c$, $\textrm{tr}(T^a T^b)=\delta_{ab}/2$, and the adjoint generator $(T^a_{adj})^{b}_{c} = -i f^{abc}$.}
The field-theory action in YM is standard. The first term in the worldline action is also standard for a free particle.
In the second term, $\psi(\tau)$ is the color wave function on the worldline  which transforms linearly under the gauge group. The point particle couples to the gauge field via the color charge $c^a= \psi^{\dagger} T^a \psi$.
Note that the worldline action manifests the reparametrization and gauge invariance, and is the leading term with lowest number of derivatives. The action can also be derived from current conservation~\cite{Goldberger:2016iau}.
To make the notation more in line with particle physics, from now on we rescale the worldline parameter $\tau'=\tau/m$ for each worldline, and $p^\mu = m v^\mu$.

The EoM then follow from the least action principle. For the position and color of a point particle, we have
\begin{align}
\frac{d^2 x^{\mu}(\tau')}{d\tau^{'2}} &= g\,  c^{a}(\tau') p^{\nu}(\tau')\, F^{a\mu}\,_{\nu} \label{eq:EoM_YM_pp0} \\
\frac{d c^a(\tau')}{d\tau'} &= g f^{abc}\,  p^{\mu}(\tau') c^{c}(\tau') \, A^{b}_{\mu}, \label{eq:EoM_YM_pp}
\end{align}
where the gauge field is evaluated at the worldline position $x^{\mu}(\tau')$. We emphasize that $x^{\mu}(\tau')$, $p^{\mu}(\tau')$, and $c^a(\tau')$ are the full time-dependent variables and are distinct from the initial conditions used later in perturbative solutions. 
After adding a Fadeev-Popov gauge fixing term $S_\textrm{GF} = -\int d^dx (\partial^\mu A^{a}_\mu)^2/2$, the EoM for gauge field can be derived
\begin{align}
\Box A^{\mu a}(x) &= -\frac{\delta S_{int}}{\delta A^{a}_{\mu}}+g\,\sum_i \int d\tau_i'\, p^{\mu}_i(\tau'_i) c^a_i(\tau'_i) \delta^{d}(x-x_i(\tau')) \label{eq:EoM_YM_rad0} \\
\frac{\delta S_{int}}{\delta A^{a}_{\mu}} &= g f^{abc} A^{b\nu} (\partial^{\mu} A^c_\nu - 2\partial_{\nu} A^{c\mu})+g f^{abc}A^b_\mu\partial^{\nu}A^c_\nu   -g^2 f^{abe} f^{ecd} A^{b\nu} A^{c\mu} A^d_\nu.
\label{eq:EoM_YM_rad}
\end{align}
We use $S_{int}$ to denote all interactions in the field-theory action. The gauge field is also sourced by all the present point particles which we now explicitly sum over. Without the point sources, the EoM is Berends-Giele recursion relation for tree-level amplitudes~\cite{Berends:1987me}.

\paragraph{Dilaton Gravity}
The minimal action for a point particle in gravity is $-m^2\,\int d\tau'$. The form of the action is fixed by power counting and reparametrization invariance. However, in the presence of dilaton $\phi$ we can dress this action with arbitrary function of $\phi$ and still maintain reparametrization and diffeomorphism invariance.
The specific coupling we need to match the double copy of YM is
\begin{equation}
S_{pp} = -m^2\,\int d\tau' e^{\kappa \phi}
\end{equation}
where the gravitational coupling $\kappa^2 = 32\pi G_N = 1/\mPl^{d-2}$ and the dilaton gravity action is
\begin{equation}
\begin{split}
S_{\textrm{GR}} & = \int d^dx \sqrt{g} \left[ -\frac{2}{\kappa^2} R+2(d-2)\, g^{\mu\nu} D_{\mu}\phi D_{\nu}\phi \right].
\label{eq:gr_rad_action}
\end{split}
\end{equation}
which is the usual Einstein-Hilbert action plus a minimally coupled dilaton.\footnote{We use mostly minus metric through out.}

The geodesic equation of a point particle is
\begin{equation}
\begin{split}
\frac{d^2 x^{\mu}(\tau')}{d\tau'^2} = -\Gamma^{\mu}_{\rho \sigma} p^{\rho}(\tau') p^{\sigma}(\tau') + \kappa m^2 g^{\mu \rho} \partial_{\rho}\phi -\kappa  p^{\mu}(\tau')p^{\rho}(\tau') \partial_{\rho} \phi
\label{eq:GR_EOM_pp}
\end{split}
\end{equation}
where $\Gamma^{\mu}_{\rho \sigma}$ is Christtoffel symbol and all the fields are evaluated at $x^\mu(\tau')$. Note that we have applied $g_{\mu\nu}p^\mu p^\nu = m^2$ as a conserved quantity to simplify the above equation. 
To derive the EoM for fields, the metric is expanded around flat space, $g_{\mu\nu}=\eta_{\mu\nu}+ \kappa h_{\mu\nu}$. 
From now on, the indices are raised and lowered by flat metric unless otherwise noted.
As in YM, we also need a gauge-fixing term $S_\textrm{GF}=\int d^d x\, (\partial_{\mu}h^{\mu}_\rho)(\partial_{\nu} h^{\nu\rho})$. The equation of motion of the graviton then follows
\begin{equation}
\Box h_{\mu\nu} = -\frac{\kappa}{2}\, \left[ \tilde{T}_{\mu\nu} - \frac{1}{d-2}\eta_{\mu\nu} \tilde{T}^{\sigma}_{\sigma} \right]
\label{eq:GR_EOM_rad1}
\end{equation}
where $\Box$ is the d'Ambertian in flat space and we introduce a coordinate dependent pseudo-tensor
\begin{equation}
\tilde{T}^{\mu\nu}(x) = -\frac{2}{\kappa}\,\frac{\delta S_{\textrm{int}}}{\delta h_{\mu\nu}} + \sum_i \int d\tau'_i\, e^{\kappa \phi(x(\tau'))}\, p^\mu_i(\tau'_i) p^\nu_i(\tau'_i) \delta(x-x_i(\tau')).
\label{eq:GR_Tmunu}
\end{equation}
as the collection of all interacting terms in EoM. The first term corresponds to the energy-momentum in spacetime, which can be expanded into an infinite tower of vertices comprised of gravitons and dilatons, and the second term is the energy-momentum of the point particles.
It is conserved, $\partial_{\mu}\tilde{T}^{\mu\nu}=0$, and can be interpreted as the tadpole term in the background field gauge effective action~\cite{Goldberger:2004jt}.
Similarly, the full EoM for dilaton reads
\begin{equation}
\sqrt{g}g^{\mu\nu} D_{\mu}\,D_{\nu}\phi = -\frac{\kappa}{4  (d-2)} \sum_i m_i \int d\tau'_i e^{\kappa \phi} \delta(x-x_i(\tau_i)).
\label{eq:GR_EOM_rad2}
\end{equation}
Note that the left hand side of the above contains the $\Box \phi$ in flat space and a tower of dilaton-graviton interactions, which will be moved to the right-hand side when solving the EoM perturbatively.

Although the procedure to find EoM seems straightforward, in practice it is daunting to expand the Einstein-Hilbert action and even the leading-order three-point vertex is barely tractable by hand, not to mention that this is only at the level of the EoM.\footnote{See Ref.~\cite{Bern:1999ji,Cheung:2016say,Cheung:2017kzx} for attempts to simplify the gravity action using insights from the double copy and amplitudes.}
In contrast, the \emph{entire} EoM for YM are summarized in Eq.~\eqref{eq:EoM_YM_pp0}, \eqref{eq:EoM_YM_pp}, \eqref{eq:EoM_YM_rad0}, and \eqref{eq:EoM_YM_rad}. The potential advantage of the double-copy construction should be clear hereafter.

\paragraph{Bi-adjoint Scalar Theory}
The bi-adjoint theory is motivated as a cousin of YM which has two copies of global color symmetries. 
The bi-adjoint scalar $\phi^{a\tilde{a}}$ carries the adjoint indices $a$ and $\tilde{a}$ for the color and the dual color group respectively. The point sources are charged under both symmetries and interact via the bi-adjoint scalar. The action is the sum of
\begin{equation}
\begin{split}
S_{pp} & = -m^2\int d\tau' + \int d\tau'\, \left( \psi^{\dagger} ip\cdot \partial\psi + y\phi^{a\tilde{a}} c^a \tilde{c}^{\tilde{a}} \right)\\
S_{\textrm{BS}} & = \int d^dx \left( \frac{1}{2} (\partial_{\mu} \phi^{a\tilde{a}})^2 -\frac{y}{3} f^{abc} \tilde{f}^{\tilde{a}\tilde{b}\tilde{c}} \phi_{a\tilde{a}} \phi_{b\tilde{b}} \phi_{c\tilde{c}} \right)
\end{split}
\end{equation}
where $y$ is a coupling constant~\cite{Goldberger:2017frp}.
All quantities in dual color are denoted with tildes. The wave function $\psi$ transform under some representation of both groups. The color and dual charge are then the contraction with respect to the respective adjoint generator, $c^a= \psi^{\dagger} \left(T^a \otimes \tilde{I}\right) \psi$ and $\tc^a= \psi^{\dagger} \left(I \otimes \tilde{T}^{\tilde{a}} \right) \psi$.
The cubic interaction of bi-adjoint scalar manifests the double-copy structure between color and the dual color group. The worldline action is then fixed by symmetries and the mass dimension of the coupling $y$. The full EoM are simply
\begin{align}
\frac{d^2 x^{\mu}(\tau')}{d\tau^{'2} } &= -y c_{a}(\tau') c_{\tilde{a}}(\tau')\partial^{\mu}\phi^{a \tilde{a}}
\label{eq:EOM_BS_pp} \\
\frac{d c^{a}(\tau')}{d\tau' } &= y f^{abc}\, c^{b}(\tau')  \tc^{\tilde{a}}(\tau')  \phi^{c \tilde{a}} \\
\frac{d \tc^{\tilde{a}}(\tau')}{d\tau' } &= y \tilde{f}^{\tilde{a}\tilde{b}\tilde{c}}\, \tc^{\tilde{b}}(\tau') c^{a}(\tau') \phi^{a \tilde{c}} 
\\
\Box \phi^{a\tilde{a}}(x) &= -yf^{abc}\tilde{f}^{\tilde{a}\tilde{b}\tilde{c}}\, \phi^{b\tilde{b}}(x) \phi^{c\tilde{c}}(x) + y\,\sum_i \,\int d\tau'_i\, c^a_i(\tau'_i) \tc^{\tilde{a}}_i(\tau'_i)\,\delta^d(x-x_i(\tau'_i)).
\label{eq:EOM_BS_rad}
\end{align} 
where all scalar fields in worldline EoM are evaluated at $x^{\mu}(\tau')$ as before.

\subsection{Perturbation and Observables}
\label{sec:perturbation}
The EoM derived in previous section are general. In this paper we solve these EoM perturbatively around a small coupling constant.
The zeroth order solution corresponds to a free particle with straight line trajectory and stationary (dual) color charge, when applicable. 
All radiation fields are turned off at zeroth order.
We parametrize the worldline degrees of freedom as
\begin{align}
x^{\mu}_i(\tau'_i) &= b^{\mu}_i + p^{\mu}_i \tau' +\delta x^{\mu}_i\\
c^{a}_i(\tau'_i) &= c^a_i + \delta c^{a}_i
\end{align}
and similarly for the dual color charge. Here $b^\mu_i$, $p^{\mu}_i$, and $c^a_i$ are the initial conditions at $\tau' \rightarrow -\infty$. The perturbation is then encoded in $\delta x^{\mu}_i$ and $\delta c^{a}_i$, whose time dependence are not shown, and the radiation field as a series expansion in coupling constants.
This perturbation regime crucially set the zeroth order as free particles with straight line trajectories, corresponding to the post-Minkowskian regime of gravity. As we emphasize in the introduction, this should be distinct from the non-relativistic perturbation of bound states, e.g., the post-Newtonian regime of binary systems.
Nevertheless, these solutions are still strictly classical and readily applicable to general relativity.

\begin{figure}[t]
	\centering
	\begin{subfigure}[t]{0.4\textwidth}
	\includegraphics[width=\textwidth, trim={4cm 17cm 5cm 2.5cm},clip,scale=.3]{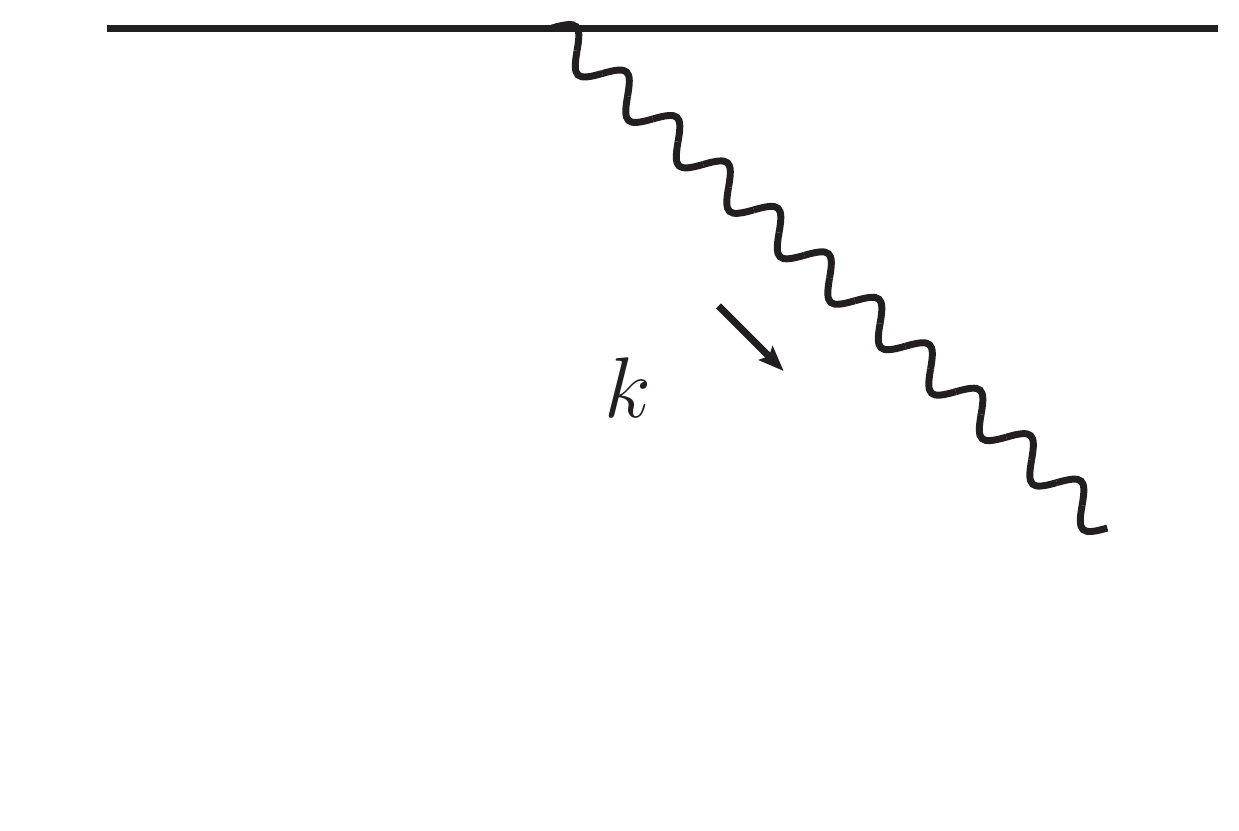}
	\caption{Radiation}
	\label{fig:single_emission}
	\hspace{5cm}
	\end{subfigure}
	~\qquad
	\begin{subfigure}[t]{0.4\textwidth}
	\includegraphics[width=\textwidth, trim={4cm 17cm 5cm 2.5cm},clip,scale=.3]{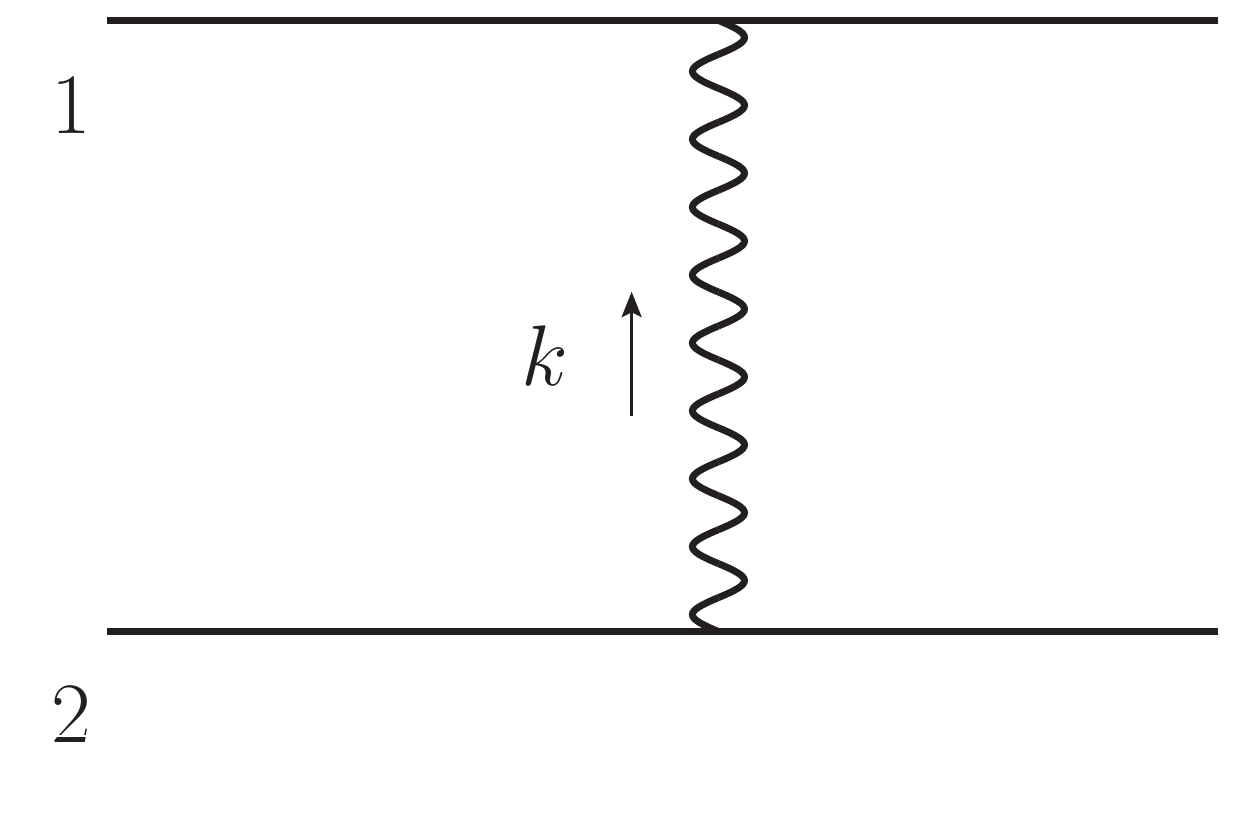}
	\caption{Deviation in worldline}
	\label{fig:LO_worldline}
	\end{subfigure}
\caption{The diagram on the right is for the radiation in Eq.~\eqref{eq:preLO}, and on the left is the diagram for worldline trajectory and color charge in Eq.~\eqref{eq:pos_LO} and Eq.~\eqref{eq:color_LO} respectively.
The solid straight lines are massive sources and the wavy lines are radiations.
}
\label{fig:LO}
\end{figure}

With the zeroth order sources, we can plug into the radiation EoM to find the fields at next order.
For example, Eq.~\eqref{eq:EoM_YM_rad} yields the gauge field emitted from the zeroth order charge $c^a$ and momentum $p^\mu$ of a particle
\begin{align}
	A^{\mu}_a (k) &= -\frac{g}{k^2} \int d\tau' \, e^{i k\cdot (b+p\tau')} \, c^a p^{\mu} \nonumber\\
		&= -\frac{g}{k^2}\, 2\pi \delta(k\cdot p)\,e^{i k\cdot b} \, c^a p^{\mu},
	\label{eq:preLO}
\end{align}
as illustrated in Figure~\ref{fig:single_emission}.
This is not yet the physical radiation at leading order for reasons we will explain soon.
This can then be plugged into into Eq.~\eqref{eq:EoM_YM_pp0} and \eqref{eq:EoM_YM_pp} to yield the deviation from straight trajectory and static color charge.


With the radiation in momentum space, the position equation of motion in Eq.~\eqref{eq:EoM_YM_pp0} at leading order becomes
\begin{equation}
\frac{d^2 \delta x^{\mu}}{d\tau^{'2}} = g\,\int \frac{d^d k}{(2\pi)^{d}}\,  e^{-ik\cdot (b+p\tau')}\,c^{a} p^{\nu}\, F^{a\mu}\,_{\nu}(k)
\end{equation}
with all higher order terms neglected.\footnote{Note that the radiation could come from a different particle and proper labeling is required.}
We can find the trajectory deviation by integrating the above twice from past infinity to the present time.
This can easily be done because the only time dependence comes from the phase of free particle trajectory. With the boundary condition imposed by $k\cdot p \rightarrow (k\cdot p+i\epsilon)$, this yields trajectory deviation at leading order
\begin{equation}
\delta x^{\mu} = -g\,\int \frac{d^d k}{(2\pi)^{d}}\,  e^{-ik\cdot (b+p\tau')} \frac{1}{(k\cdot p)^2}\,c^{a} p^{\nu}\, F^{a\mu}\,_{\nu}(k)
\label{eq:pos_LO}
\end{equation}
by plugging the field strength in Eq.~\eqref{eq:preLO}. The corresponding diagram is shown in as Figure~\ref{fig:LO_worldline}.
The time dependence of $\delta x^{\mu}$ entirely resides in the phase $e^{-ik\cdot (b+p\tau')}$.
The convenience of scattering processes is that solving the EoM is equivalent to dressing the ``propagator'' $-1/(ik\cdot p)$. The position EoM is second order in time so there is a double propagator. For color charge there is only a single propagator because its EoM is first order,
\begin{equation}
	\delta c^{a} = g\,\int \frac{d^d k}{(2\pi)^{d}}\,  e^{-ik\cdot (b+p\tau')} \frac{1}{(ik\cdot b)}\,f^{abc}c^{b} p^{\nu}\, A^{c}_{\nu}(k).
	\label{eq:color_LO}
\end{equation}
The form of this propagator is indeed related to the propagator $1/((p+k)^2-m^2)$ in field theory by taking the classical limit~\cite{Luna:2017dtq}.
The presence of both double and single propagator is one of the challenges to make the double copy work directly in the classical limit.

Let us pause for a moment and examine the small parameters used for perturbative expansions in these theories.
There are several relevant quantities in the system: the typical size of impact parameters $b$, the energy of the particles $E\gtrsim m$, the color charges $c$ and $\tc$, and finally the coupling constants. The typical scale of the radiation momentum is then set by $1/b$.
Applying the power counting to Eq.~\eqref{eq:pos_LO}, we can see the deflection $\delta x^{\mu} \ll b$ if
\begin{align}
\BS &:\quad y^2\, \frac{c^2 \tc^2}{E^3 b^{d-3}} \ll 1\\
\YM &:\quad g^2\, \frac{c^2}{E b^{d-3}} \ll 1 \\
\GR &:\quad \kappa^2\,\frac{E}{b^{d-3}} \ll 1
\end{align}
for YM and similarly other theories~\cite{Goldberger:2016iau,Goldberger:2017frp}. The phase space volume $\Gamma(\frac{d-3}{2})/(4\pi)^{\frac{d-2}{2}}$ is neglected from the above formulae.
The deviation on color charge is also under control if $c\sim \tc \sim Eb$. 
Finally, we demand the angular momentum $Eb \gg \hbar$ such that the quantum effect, namely pure graviton loop, does not enter.
Overall this implies the impact parameter should be much larger than Schwarschild radii of the sources.

We can calculate higher order results by iteration. The differences from the above are that the fields can self-interact, and the sources are not static anymore.
The latter not only corresponds to substituting the initial momentum and color charges into higher order perturbations, but an extra phase from trajectory deviation.
As we have seen the radiation is naturally solved in momentum space, as in Eq.~\eqref{eq:preLO}, while the point particles naturally live in real space.
Therefore we have to Fourier transform back and forth between the worldline and radiation. To see this, the part of YM radiation sourced by a worldline
\begin{equation}
A^{\mu}_a (k) \supset -\frac{g}{k^2} \int d\tau' \, e^{i k\cdot (b+p\tau')} \, e^{i k\cdot \delta x} \, c^a(\tau') p^{\mu}(\tau') 
\label{eq:rad_general}
\end{equation}
and the position EoM
\begin{equation}
\frac{d^2 \delta x^{\mu}}{d\tau^{'2}} = g\,\int \frac{d^d k}{(2\pi)^{d}}\,  e^{-ik\cdot (b+p\tau')} \,e^{-ik\cdot \delta x}\,c^{a}(\tau') p^{\nu}(\tau')\, F^{a\mu}\,_{\nu}(k).
\label{eq:pos_general}
\end{equation}
contain the phase $e^{\pm ik\cdot \delta x_i}$ from the path correction which is then expanded order by order. The calculation would not be consistent without this phase included. This phase makes the locality highly scrambled particularly at higher orders. For instance,
at next-to-leading order we have to include $e^{-ik\cdot \delta x} \propto (-ik\cdot \delta x)^2$ with $\delta x$ being the leading-order path deviation. 
The leading-order path deviation contains the propagator $\delta x \propto 1/(l_j \cdot p)^2$ where $l_j$ is the momentum of radiation inducing the path change. The propagator structure of $(-ik\cdot \delta x)^2$ is then $1/(l_j \cdot p)^2(l_i \cdot p)^2$ from radiation sourced by particle $i$ and $j$. 
It is difficult to assign a Feynman diagram to reproduce such pole structure along the line of Ref.~\cite{Luna:2017dtq}. Therefore we will calculate the radiation via algebraically solving the equations of motion from now on.

Nevertheless, the worldline formalism is well-established and we can push to higher orders directly.
It is helpful to streamline the structures of kinematic constraints beforehand. The radiation with momentum $k$ is a series expansion in powers of coupling
\begin{align}
A^{\mu a}(k) &= \sum_n\left(
-\frac{g^{2n+1}}{k^2}\sum_{\textrm{perm}}\, \int_1\int_2\cdots \int_{n+1} \quad (2\pi)^{d}\delta^d(k-\sum^{n+1}_{i=1} l_i)\, \mathcal{J}^{\mu a}(n) \right)
\label{eq:master_rad_YM}
\end{align}
where the phase space integration of worldline $i$ is defined as
\begin{align}
	\int_i & = \int \frac{d^{d}l_i}{(2\pi)^{d}}\, 2\pi \delta(p_i\cdot l_i) \,e^{il_i b_i}
	\label{eq:phase_space}
\end{align}
This summarizes the pattern in Eq.~\eqref{eq:preLO} as $n=0$, the leading-order results in Ref.~\cite{Goldberger:2016iau,Goldberger:2017frp,Goldberger:2017vcg,Goldberger:2017ogt,Chester:2017vcz,Li:2018qap} as $n=1$, and the next-to-leading-order results in this paper as $n=2$.
Let us go through each element in turn.
At the order $\mathcal{O}(g^{2n+1})$, the sources inserted are labeled by $i$ ranging from 1 to $n+1$. 
Each source can absorb or emit radiation as we include the back reaction. 
The corresponding phase space is integrated with the phase shift and kinematic constraint $\delta(p_i\cdot l_i)$ in Eq.~\eqref{eq:phase_space}.
The phase shift and kinematic constraint is universal for the following reasons.
As we see in Eq.~\eqref{eq:pos_general} and likewise for color charge, the absorption of momentum $l_i$ will induce a phase $e^{-il_i\cdot (b_i+p_i\tau'_{i})}$. On the other hand, the emission in Eq.~\eqref{eq:rad_general} also gives rise to a phase of opposite sign.
So if $l_i$ is the net momentum emitted by a worldline $i$, overall the time-independent part of phase shift is $e^{il_i\cdot b_i}$. 
Moreover, the time-dependent part of the phase shift yields the kinematic constraint $p_i\cdot l_i=0$ for each worldline,
after integrating the proper time from past infinity to future infinity.
The kinematic constraint can be easily understood as the on-shell condition 
$(p_i-l_i)^2-m^2_i=0$ of the point particle in classical limit~\cite{Luna:2017dtq}.
The overall momentum conservation relates radiation momentum $k$ to the total net momentum emission by worldlines $\sum_i l_i$.
Finally, if there are more than $n$ sources in the system, we have to sum over permuting the labeling of $n$ external sources into other possible combinations.
For instance, there are two source insertions for $n=1$. The permutation sum then corresponds to the sum over all possible pairs of particles in the system.

Analogous to radiation, we can also parametrize the degrees of freedom on a worldline (labeled with subscript $1$) as a sum over perturbation in coupling
\begin{equation}
\begin{split}
\delta x^{\mu}_1 &= \sum_n \, \left(
ig^{2n}\,\sum_{\textrm{perm}}\, \int_2\cdots \int_{n}\,\, e^{-i\,l_{\textrm{tot}}\cdot (b_1+p_1 \tau')} \,\delta\mathcal{X}^{\mu}_1(n) \right)\\
\delta c^{a}_1 &= \sum_n \, \left(
g^{2n}\,\sum_{\textrm{perm}}\, \int_2\cdots \int_{n}\, e^{-i\,l_{\textrm{tot}}\cdot (b_1+p_1 \tau')} \,\delta\mathcal{C}^{a}_1(n) \right).
\end{split}
\label{eq:master_worldline_ym}
\end{equation}
For order $\mathcal{O}(g^{2n})$, the worldline deviation is induced by sources from $i=2,3,\cdots n$ and $l_{\textrm{tot}}=\sum^n_{i=2}l_i$ is the total radiation momentum absorbed by the worldline. If there are more than $n$ particles in the system, the permutation sums over all the possible source insertions.
Note the time dependences of both functions entirely lie in the phase $e^{-i\,l_{\textrm{tot}}\cdot (b_1+p_1\tau')}$. 


We are only interested in the observables measured by an apparatus at spatial infinity.
The observable can be extracted by imposing retarded boundary condition, $1/k^2 = 1/((k_0+i\epsilon)^2-\vec{k}^2)$. 
To propagate the field to infinity, it amounts to localize the field at pole $k^2=0$ when integrating out the radiation momentum $k$~\cite{Porto:2016pyg}.
Let us consider the gauge field measured by a distant observer in four dimension. 
Similar formulae apply to other type of fields in general dimension. 
For an observer at distance $r$ and direction $\vec{n}$, we have
\begin{align}
A^{\pm a}(t,\vec{n}) &= \frac{1}{4\pi r} \int \frac{d\omega}{2\pi} e^{-i\omega t_r}\, \lim\limits_{k^2\rightarrow 0}\left[ -k^2\, \left( e^{\pm}_{\mu}\cdot A^{\mu a}(k) \right) \right]
\end{align}
where $t_r$ is the retarded time and $k^{\mu}=(\omega,\vec{k})$ is the momentum 4-vector with spacial direction aligned with $\vec{n}$. 
Taking the residue imposes the on-shell condition $k^2=0$ shown above.
This explains why the gauge field in Eq.~\eqref{eq:preLO} is not yet the leading-order radiation, because the on-shell condition does not have support on the kinematic constraint $\delta(k\cdot p)$.
We also need to project into the two helicity states in four dimension via a polarization vector, which satisfies the transverse condition $k\cdot e_{\pm} = 0$ and normalized, $e_{\pm}\cdot e_{\pm}^* =-1$.
It is easy calculate the energy momentum and angular momentum transmitted by the radiation once we have the above.
So the final radiation is only relevant up to the on-shell and transverse conditions.

In sum, the radiation is given order by order as Eq.~\eqref{eq:master_rad_YM}.
The explicit formulae for BS and dilaton gravity are provided in Eq.~\eqref{eq:master_bs} and Eq.~\eqref{eq:master_gr}.
The off-shell currents are rational function of the initial conditions and momentum emitted by each worldline insertion, analogous to the loop integrand in S-matrix. 
Given off-shell currents with momentum $k$, the emission amplitudes read
\begin{equation}
\begin{split}
	A_{\BS}(n) &= \lim\limits_{k^2\rightarrow 0}\,\mathcal{J}^{a\tilde{a}}(n)  \\
	A_{\YM}(n) &= \lim\limits_{k^2\rightarrow 0}\,e_{\mu}\cdot \mathcal{J}^{\mu a}(n) \\
	A_{h}(n) &=  \lim\limits_{k^2\rightarrow 0}\,e_{\mu\nu}\cdot \mathcal{J}^{\mu \nu}(n) \\
	A_{\phi}(n) &= \lim\limits_{k^2\rightarrow 0}\,\mathcal{J}(n)
	\label{eq:amp_def}
\end{split}
\end{equation}
with the on-shell condition imposed.
The (dual) color indices are omitted in the emission amplitudes.
In generic dimension, the polarization vector for YM satisfies $k\cdot e = 0$.
For gravity, the polarization is symmetric, transverse, and traceless, $k^\mu e_{\mu\nu} =e^{\mu}_{\mu}= 0$.
These on-shell amplitudes are gauge-invariant objects where the double copy applies.


\subsection{Color and Kinematics}
\label{sec:bcj_review}
Let us first review the color-kinematics duality and double copy in field theory.
A tree-level amplitude in Yang-Mills can be expressed as
\begin{equation}
\begin{split}
\sum_{i} \frac{C_i N_i}{D_i}
\label{eq:YM_qft}
\end{split}
\end{equation}
where $C_i$ and $N_i$ is the respective color and kinematic numerator, and the propagator denominator $D_i$ associated with cubic diagram $i$. Note that the color and propagators are one-to-one to a cubic diagram which will not be the case for classical radiation. 
There is an intrinsic ambiguity in $N_i$ because the Yang-Mills theory has a quartic vertex. 
The Yang-Mills amplitude is the gauge invariant under the sum over different diagrams, which are related by color identities in Eq.~\eqref{eq:color_id_qft}.
Crucially the gauge invariance only relies on these algebraic relations, regardless of details in color factors.

A remarkable observation by Bern, Carrasco, and Johansson is that the double copy relation is manifest through the color-kinematics duality in Eq.~\eqref{eq:kin_id_qft}. If we replace color factors with kinematic numerators
\begin{equation}
C_i \, \rightarrow \, N_i,
\label{eq:bcj}
\end{equation}
then a YM amplitude in Eq.~\eqref{eq:YM_qft} becomes
\begin{equation}
\begin{split}
	\sum_{i} \frac{N_i N_i}{D_i},
	\label{eq:YMSquare_qft}
\end{split}
\end{equation}
which turns out to be an amplitude in gravity!
This miraculous connection is based on two pillars---locality and gauge invariance. The locality structure, encoded in the propagators, is preserved during the double copy construction.\footnote{More generally, double-copy construction can still hold even with non-local numerators.}
Second, the gauge invariance relies on the color identities, which are obeyed by the kinematic numerators in Eq.~\eqref{eq:kin_id_qft}. 
Therefore the gauge invariance is not only preserved in Eq.~\eqref{eq:YMSquare_qft} as the original YM, but is enhanced to (linearized) diffeomorphism invariance in gravity via double-copy structure.
Being both local and gauge invariant, the double copy is justified as a gravitational amplitude which can be proven rigorously~\cite{Bern:2010yg,Arkani-Hamed:2016rak}.

The success of double copy hinges on the color-kinematics duality. However, it is far from obvious that such duality-satisfying representation exists. In particular, the Jacobi identity imposes non-trivial constraints on the numerators. The existence of such representation, the so-called BCJ numerators, has been proven at tree level but remains to be a conjecture beyond this. Even the existence is known, finding explicit expression for BCJ numerators requires non-trivial effort.
See Ref~\cite{Carrasco:2015iwa,Cheung:2017pzi} for more details.

Many features of double copy can be learned by studying the bi-adjoint scalar amplitude
\begin{equation}
	\begin{split}
	\sum_{i} \frac{C_i \tilde{C}_i}{D_i}
	\label{eq:BS_qft}
	\end{split}
\end{equation}
where we replace the kinematic numerators in YM with dual color factors~\cite{Carrasco:2015iwa,Cheung:2017pzi}. 
The bi-adjoint scalar theory shares the same traits as any other theories with color-kinematics duality. For instance, the color-stripped amplitudes, which are the amplitudes when we project Eq.~\eqref{eq:BS_qft} to a minimal basis of color factors, are still dependent under the so-called BCJ amplitude relations. 
When we stripped both color factors and remove the redundancies from amplitude relations, these doubly stripped amplitudes form a matrix whose inverse becomes the KLT kernel. The very same kernel allows us to double copy from YM to gravity directly at the level of amplitudes.
We will see that bi-adjoint scalar theory plays an important role in worldline formalism as well.

The gravity amplitude obtained from YM double copy is not just Einstein gravity, but contains other fields called factorized gravity.\footnote{The name is not yet settled in literatures. It is also called fat gravity, extended gravity, or $\mathcal{N}=0$ supergravity.}
From degrees of freedom counting in little group, it is obvious that the YM double copy contains more than the usual graviton. Specifically, if we use the polarization $e_{\mu}$ and $\tilde{e}_{\nu}$ for each of the YM copy, the resulting polarization $e_{\mu}\tilde{e}_{\nu}$ in gravity is factorized which can be decomposed as three terms
\begin{align}
e_{\mu}\tilde{e}_{\nu} &= \frac{1}{2}\left(e_{\mu}\tilde{e}_{\nu}+\tilde{e}_{\mu}e_{\nu} - 2\,\frac{\,e\cdot \tilde{e}}{d-2}P_{\mu\nu} \right) 
+ \frac{1}{2}\left( e_{\mu}\tilde{e}_{\nu}-\tilde{e}_{\mu}e_{\nu} \right)
+ \frac{\,e\cdot \tilde{e}}{d-2}P_{\mu\nu},
\label{eq:decomposition}
\end{align}
where $P_{\mu\nu} = \eta_{\mu\nu}-\left(k_{\mu}q_{\nu}+q_{\mu}k_{\nu}\right)/\inn{k}{q}$ is a projector transverse to momentum $k_{\mu}$ and a light-like reference
$q_{\mu}$.\footnote{The final results are independent of the reference vector.}
The first term is symmetric and traceless, and thus corresponds to the usual graviton. The second one corresponds to an anti-symmetric two-form, and the trace part from the last term gives a dilaton. If the two-form is not present, we have a theory with graviton and dilaton whose action is given by Eq.~\eqref{eq:gr_rad_action}. 
Although it is possible to double copy gauge theories into Einstein gravity, as demonstrated~\cite{Johansson:2014zca,Luna:2017dtq}, we only consider dilaton gravity in this work as an initial study at next-to-leading order.

\section{Summary of Calculation}
\label{sec:calculation}
\begin{figure}[t]
	\centering
	\includegraphics[trim={0 15cm 0 0},clip,scale=.4]{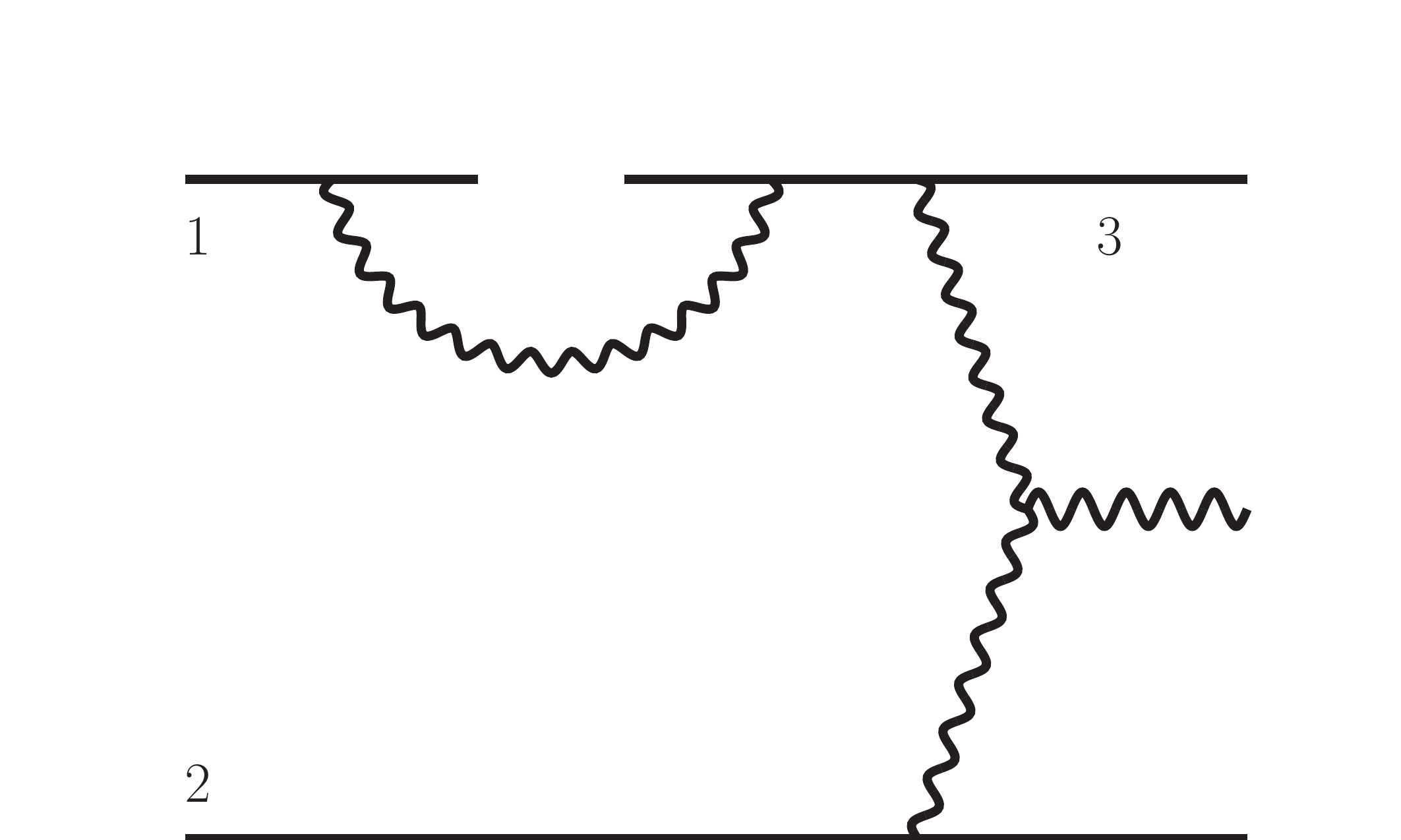}
	\caption{A schematic diagram contributed to next-to-leading order radiation in the same notation as Figure~\ref{fig:LO}.
		This becomes a self-energy correction to particle 1 with leading order radiation if we identify the particle 3 with particle 1.}
	\label{fig:self}
\end{figure}

We now proceed to calculate radiation at higher order in various theories. 
At the next-to-leading order, corresponding to fifth power in coupling constant,
there are three source insertions for radiation with $n=1,2,3$. 
Without loss of generality, we can first treat the third insertion as an independent third body, with its own initial momentum and color charge, and calculate the current involving three sources. 
If we want to retrieve the classical radiation in a two-body system, the current is then extracted from a generic three-body system by identifying the initial conditions of the sources. 
Take YM current as an example,
\begin{equation}
\mathcal{J}^{\mu a}(\textrm{2-body}) = \frac{1}{2!} \left(\mathcal{J}^{\mu a}(\textrm{3-body})\big\vert_{p_3=p_1,c_3=c_1} + \mathcal{J}^{\mu a}(\textrm{3-body})\big\vert_{p_3=p_2,c_3=c_2} \right)
\label{eq:3_body}
\end{equation}
where we set the initial momentum $p_3$ as $p_1$ or $p_2$ and likewise for the charges. Note the net momentum $l_3$ emitted by the third source insertion should remain to be independent to $l_{1,2}$.
Since the three-body current assumes the sources are distinct, there is an overall symmetry factor $2!$ for the over-counting of identifying two sources.
Similar formulae hold for the trajectory and color charge at next-to-leading order.
These integrands are analogous to the bare integrand in field theories, including self-energy renormalization to leading order. See Figure \ref{fig:self} for an illustration.
Although this is book-keeping method, it is important for seeing the double-copy structure as we will discuss in Sec.~\ref{sec:discussion}. 
Unless otherwise noted, from now on we consider a three-body system with initial conditions $p_{1,2,3}$, and the (dual) color charges $c_{1,2,3}$ and $\tc_{1,2,3}$ when applicable.
This avoids extra terms in permutation sum in Eq.~\eqref{eq:master_rad_YM} at next-to-leading order.

\subsection{Bi-adjoint Scalar Theory}
\label{sec:bs}
The perturbative solutions of bi-adjoint theory follow from EoM in Eq.~\eqref{eq:EOM_BS_pp} to Eq.~\eqref{eq:EOM_BS_rad}.
Analogous to YM, we also factor out all kinematic constraints and phase space integration
\begin{equation}
\begin{split}
\phi^{a \tilde{a}}(k) &= \sum_n\left(
-\frac{y^{2n+1}}{k^2}\sum_{\textrm{perm}}\, \prod^{n+1}_{i=1}\,\int_i\quad (2\pi)^{d}\delta^d(k-\sum l_i)\, \mathcal{J}^{a\tilde{a}}(n) \right) \\
\delta x^{\mu}_1 &= \sum_{n}\left( iy^{2n}\,\sum_{\textrm{perm}}\, \prod^n_{i=1}\,\int_i \, e^{-i\,l_{\textrm{tot}}\cdot (b+p\tau')}\, \delta\mathcal{X}^{\mu}_1 (n)
\right)
\\
\delta c^{a}_1 &= \sum_n\left( 
y^{2n}\,\sum_{\textrm{perm}}\, \prod^n_{i=1}\,\int_i\, e^{-i\,l_{\textrm{tot}}\cdot (b+p\tau')}\, \delta\mathcal{C}^{a}_1(n)
\right)
\end{split}
\label{eq:master_bs}
\end{equation}
for the bi-adjoint current and worldline degrees of freedom, following notations in Eq.~\eqref{eq:master_rad_YM} and Eq.~\eqref{eq:master_worldline_ym}. The expression for dual color can be found straightforwardly.

\paragraph{Leading Order} Let us first summarize the results at leading order from Ref.~\cite{Goldberger:2017frp}.
To linear order in coupling ($n=0$), the bi-adjoint scalar current emitted by particle $2$ with momentum $l_2$ is
\begin{equation}
	\JBS(0) = \frac{c^a_2 \,\tc^{\tilde{a}}_2}{l^2_2}
\end{equation}
using the notation in Eq.~\eqref{eq:master_bs} and the subscript labels the particle.
After this radiation is absorbed by particle $1$, its trajectory and color deviation are
\begin{align}
	\delta\mathcal{X}^{\mu}_1(1) &= \frac{(\inn{c_1}{c_2}) (\inn{\tc_1}{\tc_2}) l^\mu_2}{l^2_2 (\inn{p_1}{l_2})^2} \\
	\delta\mathcal{C}^{a}_1(1) &= - \frac{(\inn{\tc_1}{\tc_2}) \left[c_1,c_2\right]^{a}}{l^2_2 (\inn{p_1}{l_2})},
\end{align}
where $[c_1,c_2]^a \equiv i f^{abc} c^b_1 c^c_2$ and similarly the deviation of particle $2$.
Since there is only one source inserted at this level, we omit the trivial sum in Eq.~\eqref{eq:master_bs} as can be done by permutations.

To present the leading-order radiation, we introduce the notion of color stripping. To do so, we first reduce the color factors to a minimal basis, \begin{equation}
	\JBS(n) = \sum_{i}\, C_i\, \JBS(n,C_i).
	\label{eq:color_stripping_LO}
\end{equation}
The function $\JBS(n,C_i)$ associated with each color factor $C_i$ is the color-stripped object,
and likewise for the emission amplitude and point source.\footnote{We abuse the notation slight by using the same $C_i$ for color factors in a minimal basis but it should be clear from the context.}
At leading order, the independent color factors of radiation emitted by particle $1,2$ are
\begin{equation}
C_1 = (c_ 1\cdot c_ 2) c_1^a,\quad C_2=(c_ 1\cdot c_ 2) c_2^a, \quad C_3=[c_1 ,c_2]^a
\label{eq:g3_color}
\end{equation}
The explicit color-stripped currents are
\begin{equation}
\begin{split}
	\JBS(1,C_1) =&  
	-\tC_1 \,\frac{l_2\cdot l_{12}  }{l^2_2 \left(l_2\cdot p_1\right)^2}
	-\tC_3 \,\frac{1}{l^2_2 ({l_2\cdot p_1})} \\
	\JBS(1,C_2) =& 
	-\tC_2 \,\frac{l_1\cdot l_{12}  }{l^2_1 \left(l_1\cdot p_2\right)^2}
	+\tC_3 \,\frac{1}{l^2_1 ({l_1\cdot p_2})} \\
	\JBS(1,C_3) =& 
	-
	\tC_1 \frac{1}{l^2_2 (l_2\cdot p_1)}+\tC_2 \frac{1}{l^2_1 (l_1\cdot p_2)}+\tC_3 \frac{2}{l^2_1 l^2_2}
	\label{eq:bs_rad_LO}
\end{split}
\end{equation}
with the momentum of the radiation $l_{12}=l_{1}+l_2$.
The off-shell currents serve as inputs for higher order radiation.

From the above results, we can see two exotic features in classical radiation from field-theory amplitudes. First, there are non-trivial numerators even for bi-adjoint scalar theory. Second, the color and dual color is no longer diagonal. Both are features absent in field-theory amplitude and clarify the formulation of double copy in classical radiation.

\paragraph{Next-to-leading Order} 
Using the inputs from leading order and EoM, we find the next-to-leading-order worldline trajectory and (dual) color charge. The results are summarized in Appendix~\ref{app:worldline}. For the radiation, the full list of numerators are given in Table~\ref{table:g5_color}. We can decompose the current by these color numerators with $C_{21}$ removed via Jacobi identity. The first color-stripped current
\begin{align}
	\JBS_{\BS}(2,C_1) =& 
	\frac{\tC_1}{l^2_2 l^2_3}\left(
		\frac{(l_{123}\cdot l_2) (l_{123}\cdot l_3)}{ \left(l_2\cdot p_1\right)^2 \left(l_3\cdot p_1\right)^2}
		-\frac{(l_2\cdot l_3) (l_{123}\cdot l_3)}{ \left(l_2\cdot p_1\right)^2 \left(p_1\cdot l_{23}\right)^2}
		-\frac{(l_2\cdot l_3) (l_{123}\cdot l_2)}{ \left(l_3\cdot p_1\right)^2 \left(p_1\cdot l_{23}\right)^2} \right) \nonumber\\
	&+\frac{\tC_4}{l^2_2 l^2_3} \left( \frac{l_{123}\cdot l_3}{ (l_2\cdot p_1) \left(p_1\cdot l_{23}\right)^2}-\frac{l_{123}\cdot l_2}{ (l_3\cdot p_1) 		\left(p_1\cdot l_{23}\right)^2} \right) \nonumber\\
	&+\frac{\tC_{13}}{l^2_2 l^2_3} \left(\frac{ l_{123}\cdot l_3}{ (l_2\cdot p_1) \left(l_3\cdot p_1\right)^2}-\frac{l_2\cdot l_3}{ \left(l_3\cdot p_1\right)^2 p_1\cdot l_{23}} \right) 
	+\frac{\tC_{17}}{l^2_2 l^2_3} \left(\frac{l_2\cdot l_3}{ \left(l_2\cdot p_1\right)^2 p_1\cdot l_{23}}-\frac{l_{123}\cdot l_2}{ \left(l_2\cdot p_1\right)^2 l_3\cdot p_1}\right) \nonumber\\
	&+\frac{\tilde{C}_{20}}{l^2_2 l^2_3 (l_3\cdot p_1) (p_1\cdot l_{23})}-\frac{\tilde{C}_{21}}{l^2_2 l^2_3 (l_2\cdot p_1) (p_1\cdot l_{23})},
\label{eq:bs_rad_NLO_1}
\end{align}
is symmetric under the exchange of particle $2,3$ with $l_{ijk}=l_i+l_j+l_k$. Also,
\begin{align}
\JBS_{\BS}(2,C_4) =&
+\frac{1}{l^2_2 l^2_3 \left(p_1\cdot l_{23}\right)^2} \left(
	 \frac{l_{123}\cdot l_3}{ l_2\cdot p_1 } \,\tilde{C}_1
	-\frac{l_{123}\cdot l_2}{ l_3\cdot p_1 } \,\tilde{C}_1
	-\frac{2 l_{123}\cdot l_{23}}{l^2_{23}}  \,\tilde{C}_4 
\right) \nonumber\\
&
+\frac{1}{l^2_3 l^2_{23} (l_3\cdot p_2) \left(p_1\cdot l_{23}\right)} \left(
	 \tilde{C}_{14}+\frac{l_{123}\cdot l_{23}}{p_1\cdot l_{23}}\,\tilde{C}_7 
\right)
+\frac{1}{l^2_2 l^2_{23} (l_2\cdot p_3) \left(p_1\cdot l_{23}\right)} \left(
	 \tilde{C}_{18}-\frac{l_{123}\cdot l_{23}}{p_1\cdot l_{23}} \,\tilde{C}_8 
\right)
\nonumber\\
&
-\frac{1}{l^2_2 l^2_3 (p_1\cdot l_{23})}\left(
	 \frac{\tilde{C}_{17}}{l_2\cdot p_1}
	+\frac{\tilde{C}_{13}}{l_3\cdot p_1 }
	+\frac{2 \tilde{C}_{19}}{l^2_{23} }
\right),
\label{eq:bs_rad_NLO_2}
\end{align}
is anti-symmetric under the exchange of particle $2,3$. 
The follow two color-stripped amplitudes do not have an isometry under the exchange of particles
\begin{align}
\JBS_{\BS}(2,C_7) =& 
\frac{l_{123}\cdot l_{23}}{l^2_3 l^2_{23} \left(l_3\cdot p_2\right) \left(p_1\cdot l_{23}\right)^2}\left(
\tilde{C}_4 +\frac{l_3\cdot l_{23}}{l_3\cdot p_2}\,\tilde{C}_7
\right)
+\frac{1}{l^2_3 l^2_{23} \left(l_3\cdot p_2\right) (p_1\cdot l_{23})} \left(
\tilde{C}_{19} +\frac{l_3\cdot l_{23}}{l_3\cdot p_2}\tilde{C}_{14}
\right), \nonumber \\ \\
\JBS_{\BS}(2,C_{13}) =& 
-\frac{1}{l^2_2 l^2_3 (l_2\cdot p_1) \left(l_3\cdot p_1\right)}\left(
\tilde{C}_{17} - \frac{l_{123}\cdot l_3}{l_3\cdot p_1} \,\tilde{C}_1
\right)
-\frac{1}{l^2_2 l^2_3 \left(l_3\cdot p_1\right) (p_1\cdot l_{23})} \left(
\tilde{C}_4+\frac{l_2\cdot l_3}{l_3\cdot p_1}\tilde{C}_1
\right)
\nonumber \\
&
+\frac{1}{l^2_3 l^2_{13} \left(l_3\cdot p_1\right) (p_2\cdot l_{13})}\left(
\tilde{C}_5-\frac{l_3\cdot l_{13}}{l_3\cdot p_1}\, \tilde{C}_9
\right)
-\frac{2}{l^2_2 l^2_3 l^2_{13} \left(l_3\cdot p_1\right)} \left(
\tilde{C}_{20} +\frac{l_3\cdot l_{13}}{l_3\cdot p_1}\, \tilde{C}_{13}
\right).
\label{eq:bs_rad_NLO_3}
\end{align}
By removing $C_{21}$ via Jacobi identity, the last color-stripped amplitude is associated is
\begin{align}
\JBS_{\BS}(2,C_{19}) =& 
\,\frac{1}{l^2_2 l^2_3 (p_1\cdot l_{23})} \left(
	 \frac{\tilde{C}_1}{l_2\cdot p_1} 
	-\frac{2 \tilde{C}_4}{l^2_{23} } \right)
+\frac{1}{l^2_1 l^2_2 (p_3\cdot l_{12})}\left(
	 \frac{\tilde{C}_3}{l_2\cdot p_3}
	+\frac{2 \tilde{C}_6}{l^2_{12}} \right) 
-\frac{\tilde{C}_2}{l^2_1 l^2_3 (l_1\cdot p_2) (l_3\cdot p_2)}	
\nonumber\\
&
+\frac{1}{l^2_{23} (p_1\cdot l_{23})}
	\left(\frac{\tilde{C}_7}{l^2_3  (l_3\cdot p_2) }
	-\frac{\tilde{C}_8}{l^2_2 (l_2\cdot p_3)}\right)
+\frac{1}{l^2_{12} (p_3\cdot l_{12})}
	\left(\frac{\tilde{C}_{11} }{l^2_1  (l_1\cdot p_2)}
	-\frac{\tilde{C}_{12}}{l^2_2 (l_2\cdot p_1)} \right) \nonumber\\
&
+\frac{1}{l^2_3 l^2_{12} }\left(
	 \frac{2 \tilde{C}_{15}}{l^2_1 (l_1\cdot p_2)}
	+\frac{2 \tilde{C}_{17}}{l^2_2 (l_2\cdot p_1)}\right)
-\frac{1}{l^2_1  l^2_{23}  }\left(
	 \frac{2 \tilde{C}_{14}}{l^2_3 (l_3\cdot p_2)}
	+\frac{2 \tilde{C}_{18}}{l^2_2 (l_2\cdot p_3)}
	\right) \nonumber\\
&+\frac{1}{l^2_1 l^2_2 l^2_3}\left(
	\frac{4 \tilde{C}_{19}}{l^2_{23}}
	-\frac{4 \tilde{C}_{21}}{l^2_{12}} \right). 
\label{eq:bs_rad_NLO_4}
\end{align}
Note that it does not inherit the antisymmetry of $2\leftrightarrow 3$ in $C_{19}$, because we have used Jacobi identity.
All other color-stripped amplitudes are related from these by permutation. For example, the isometry of color numerators implies $A_{\BS}(2,C_{20}) = -A_{\BS}(2,C_{19})\vert_{1\leftrightarrow 2}$. 
An ancillary file \texttt{bs\_radiation\_NLO.m} is attached for $A_{2,\BS}$.
Unfortunately, the next-to-leading-order radiation is lengthy even for the simplest bi-adjoint scalar theory. After all, it should be compared with a seven-point tree amplitude or a five-point amplitude at one loop. The level of complexity only keeps growing for YM and gravity.

\begin{table}[t]
	\def\arraystretch{1.5}
	\centering
	\begin{tabular}{LLLL}
		\hline\hline
		C_1=(c_1\cdot c_2) (c_1\cdot c_3) c^a_1 & C_{7}\,\,=(c_1\cdot c_2) (c_2\cdot c_3) c^a_1 & C_{13} = (c_1\cdot c_3) \left[c_1,c_2\right]^a 
& C_{19}= \left[c_1,\,\left[c_2,\,c_3\right] \right]^a \\
		C_2=(c_1\cdot c_2) (c_2\cdot c_3) c^a_2 & C_{8}\,\,=(c_1\cdot c_3) (c_2\cdot c_3) c^a_1 & C_{14} = (c_2\cdot c_3) \left[c_1,c_2\right]^a 
& C_{20}= \left[c_2,\,\left[c_3,\,c_1\right] \right]^a \\
		C_3=(c_1\cdot c_3) (c_2\cdot c_3) c^a_3 & C_{9}\,\,=(c_1\cdot c_2) (c_1\cdot c_3) c^a_2 & C_{15} = (c_1\cdot c_2) \left[c_2,c_3\right]^a 
& C_{21}= \left[c_3,\,\left[c_1,\,c_2\right] \right]^a \\
		C_{4} = \left(c_1\cdot \left[c_2,c_3\right]\right)\, c^a_1 
		& C_{10}=(c_1\cdot c_3) (c_2\cdot c_3) c^a_2 & C_{16} = (c_1\cdot c_3) \left[c_2,c_3\right]^a & \\
		C_{5} = \left(c_1\cdot \left[c_2,c_3\right]\right)\, c^a_2
		& C_{11}=(c_1\cdot c_2) (c_2\cdot c_3) c^a_3 & C_{17} = (c_1\cdot c_2) \left[c_3,c_1\right]^a & \\
		C_{6} = \left(c_1\cdot \left[c_2,c_3\right]\right)\, c^a_3 
		& C_{12}=(c_1\cdot c_2) (c_1\cdot c_3) c^a_3 & C_{18} = (c_2\cdot c_3) \left[c_3,c_1\right]^a & \\
		\hline\hline
	\end{tabular}
	\caption{List of the color numerators relevant for radiation at next-to-leading order. The last three satisfy Jacobi identity $C_{19}+ C_{20} + C_{21}=0$.}
	\label{table:g5_color}
\end{table}

\subsection{Yang-Mills Theory}
The general feature of YM has been introduced in Section~\ref{sec:setups}. The EoM are given in Eq.~\eqref{eq:EoM_YM_pp0}, \eqref{eq:EoM_YM_pp}, \eqref{eq:EoM_YM_rad0}, and \eqref{eq:EoM_YM_rad}. The gauge current and the deviation of a point particle's degrees of freedom are parametrized in Eq.~\eqref{eq:master_rad_YM} and~\eqref{eq:master_worldline_ym} respectively. 

\paragraph{Leading Order}
The gauge field current to linear order in coupling is given in Eq.~\eqref{eq:preLO}. The leading-order deviation of point particle $1$ is
\begin{align}
	\delta\mathcal{X}^{\mu}_1(1) &= \frac{\inn{c_1}{c_2}}{l^2_2 (\inn{p_1}{l_2})}
	\left(p^\mu_2 -\frac{\inn{p_1}{p_2}}{\inn{p_1}{l_2}}\, l^\mu_2 \right)\\
	\delta\mathcal{C}^{a}_1(1) &= \frac{\inn{p_1}{p_2}}{l^2_2 (\inn{p_1}{l_2})} \, \left[c_1,c_2\right]^{a}.
\end{align}
Combining the self-interaction of gluon and the emission from point particles with leading-order deviation, the leading-order radiation with momentum $l_1 + l_2$ from particle $1,2$ is
\begin{equation}
	\begin{split}
		\JYM(1,C_1) =&  \frac{\inn{p_1}{p_2}}{l^2_2 (l_2\cdot p_1)}\left(\frac{l_2\cdot l_{12}}{l_2\cdot p_1}\, p^\mu_1 -l^{\mu }_2\right)
		+\frac{1}{l^2_2}\left(p^{\mu }_2- \frac{p_2\cdot l_{1}}{l_2\cdot p_1}p^{\mu }_1 \right)
		\\
		\JBS(1,C_2) =& 
		\frac{\inn{p_1}{p_2}}{l^2_1 (l_1\cdot p_2)}\left(\frac{l_1\cdot l_{12}}{l_1\cdot p_2}\, p^\mu_2 -l^{\mu }_1\right)
		+\frac{1}{l^2_1}\left(p^{\mu }_1- \frac{p_1\cdot l_{2}}{l_1\cdot p_2}p^{\mu }_2 \right)
		\\
		\JBS(1,C_3) =& 
		\frac{2}{l^2_1 l^2_2}\left( (l_1\cdot p_2) p^{\mu }_1 - (l_2\cdot p_1)p^{\mu }_2 \right)
		+\frac{p_1\cdot p_2}{l^2_1 l^2_2}\left(\frac{l^2_1}{l_2\cdot p_1}p^{\mu }_1-\frac{l^2_2}{l_1\cdot p_2}p^{\mu }_2- l^{\mu }_1+l^{\mu }_2 \right)
	\end{split}
\end{equation} 
where we use the color-stripping decomposition in Eq.~\eqref{eq:color_stripping_LO} with the color numerators in Eq.~\eqref{eq:g3_color}.
Note that each color-stripped current satisfies Ward identity, $\partial_{\mu} \JYM(n,C_i)$, even for off-shell current.

\paragraph{Next-to-Leading Order}
The trajectory and color deviation at next-to-leading order are given in \texttt{ym\_pos\_NLO.m} and \texttt{ym\_color\_NLO.m} with relevant color factors in Appendix~\ref{app:worldline}. The next-to-leading-order gluon contains color numerators in Table~\ref{table:g5_color}. The full expression of $\JYM(2)$ can be found in \texttt{ym\_radiation\_NLO.m}. We have verified Ward identity for the off-shell current.

\subsection{Dilaton Gravity}
The gravity calculation follow from EoM in Eq.~\eqref{eq:GR_EOM_pp}, Eq.~\eqref{eq:GR_EOM_rad1}, and~\eqref{eq:GR_EOM_rad2}.
Analogous to other theories, let us define~\footnote{The normalization of graviton and dilaton current are different from the bi-adjoint scalar theory and YM.}
\begin{equation}
\begin{split}
h^{\mu \nu}(k) &=\sum_n \,\left(\frac{1}{k^2}\,\left(\frac{\kappa}{2}\right)^{2n+1}\,\sum_{\textrm{perm}}\, \prod^{n+1}_{i=1}\,\int_i\quad (2\pi)^{d}\delta^d(k-\sum l_i)\, \mathcal{J}^{\mu \nu}(n) \right)\\
\phi(k) &= \sum_n \,\left( \frac{1}{k^2}\,\left(\frac{\kappa}{2}\right)^{2n+2}\,\sum_{\textrm{perm}}\, \prod^{n+1}_{i=1}\,\int_i\quad (2\pi)^{d}\delta^d(k-\sum l_i)\,  \mathcal{J}(n) \right)\\
\delta x^{\mu}_1 &= \sum_n \,\left( i\left(\frac{\kappa}{2}\right)^{2n}\,\sum_{\textrm{perm}}\, \prod^n_{i=1}\,\int_i \, e^{-i\,l_{\textrm{tot}}\cdot (b+p\tau')}\, \delta\mathcal{X}^{\mu}_1(n) \right)
\end{split}
\label{eq:master_gr}
\end{equation}
for the off-shell graviton and dilaton currents, and the trajectory of a point particle. Note the indices in $h^{\mu \nu}$ are raised  from $h_{\mu \nu}$ by flat metric.
We can also define an off-shell integrand for the pseudo stress-energy tensor in Eq.~\eqref{eq:GR_Tmunu}
\begin{equation}
\begin{split}
	T^{\mu \nu}(k) &= \sum\,\left(
	\left(\frac{\kappa}{2}\right)^{2n}\,\sum_{\textrm{perm}}\, \prod^{n+1}_{i=1}\,\int_i\quad (2\pi)^{d}\delta^d(k-\sum l_i)\, \mathcal{T}^{\mu \nu} (n)
	\right)
\end{split}
\label{eq:master_tmunu}
\end{equation}
such that 
\begin{equation}
	\mathcal{J}^{\mu \nu}(n) = \mathcal{T}^{\mu \nu}(n) -\frac{\eta^{\mu\nu}}{d-2} \mathcal{T}^{\sigma}_{\sigma}(n)
\end{equation}
A non-trivial check for the calculation is the conservation of $\partial_{\mu}\mathcal{T}^{\mu \nu}(n)=0$ on the support of the above kinematic constraints, even for off-shell momentum.
For instance, the currents at $\mathcal{O}(\kappa)$ emitted by a point particle with momentum $p$ are $\TGR(0) = p^{\mu}p^{\nu}$ and $\Jdilaton(0) = p^2/(d-2)$.
The stress-energy is not traceless for massive sources, and therefore dilaton is  produced. It is obvious that $\partial_{\mu}\mathcal{T}^{\mu \nu}(n)=0$ by the kinematic constraint in Eq.~\eqref{eq:master_gr}. The leading order results can be found in Ref.~\cite{Goldberger:2016iau} which we completely reproduce.

The trajectory deviation at $\mathcal{O}(\kappa^4)$ is also available as the function $\delta\mathcal{X}^{\mu}_1(2)$ in the ancillary file \texttt{gr\_pos\_NLO.m}. As we choose the worldline parametrization to be (proportional to) the proper time, we have
$g_{\mu\nu}(x(\tau')) p^{\mu}(\tau') p^{\nu}(\tau') = m^2$
as a constant at any time $\tau'$. We have verified this at next-to-leading order as a non-trivial check to trajectory deviation $\delta\mathcal{X}^{\mu}_1(2)$. Note that for this equality to hold, the metric has to be evaluated at the worldline trajectory $x^\mu(\tau')$ to necessary order.

For next-to-leading-order radiation, 
$\TGR(2)$ for graviton is provided in \texttt{gr\_radiation\_NLO.m} and and the current $\Jdilaton(2)$ for dilaton is given in \texttt{gr\_dilaton\_NLO.m}. First, we have verified Ward identity $\partial_{\mu}\mathcal{T}^{\mu \nu}(2)=0$ even for off-shell stress-energy tensor.
Also, the graviton emission amplitude is independent of spacetime dimension, as a feature of dilaton gravity.

\section{Color-Kinematics Duality}
\label{sec:bcj}
Now we are ready to discuss the color-kinematics duality in classical radiation. 
With the previous calculation, physical emission amplitudes are extracted via Eq.~\eqref{eq:amp_def}.
The amplitudes of classical radiation in worldline formalism are different from field theory in many places.
As we have seen concretely in Section~\ref{sec:calculation}, locality is obscured by the presence of both double and single propagators, and also the phase perturbation from trajectory deviation. 
Even for the simple bi-adjoint scalar theory in Section~\ref{sec:bs}, there are kinematic numerators in the emission amplitudes and the two copies of color factors are not diagonal. We lose the field-theory decomposition in Eq.~\eqref{eq:YM_qft}. As locality becomes opaque, 
it is not obvious how to define double copy, nor whether a duality-satisfying representation exist at all. Any proposal as a square to gravity has to be examined explicitly as there is no rigorous proof. We will also study the BCJ and KLT relations for classical radiation since these are the direct consequences of color-kinematics duality.
Let us restrict to physical emission amplitudes now and discuss these issues in turn.

\subsection{Double Copy in Worldline Formalism}
Built on the hints from leading order studies~\cite{Goldberger:2016iau,Goldberger:2017frp,Luna:2017dtq,Goldberger:2017ogt,Goldberger:2017vcg,Li:2018qap,Chester:2017vcz}, we propose a simple resolution: use the bi-adjoint scalar theory to identify those kinematic terms that are suppose to be propagators. Concretely, we write the emission amplitude in bi-adjoint scalar theory as
\begin{equation}
A_{\BS}(n) = (-1)^{n}\,\sum_{i,j} C_i\, P_{ij}\,\tilde{C}_j
\label{eq:bs_worldline_amp}
\end{equation}
where we sum over all the color numerators $C_{i},\tilde{C}_j$ and $n=0,1,2,\cdots$ corresponds to the order $\mathcal{O}(y^{1+2n})$.
The locality is encoded in $P_{ij}$, which is not necessary diagonal, and nor only lies in the denominator. Nevertheless, we use the amplitude in bi-adjoint scalar theory to define locality, and span the YM amplitude accordingly
\begin{equation}
A_{\YM}(n) = \sum_{i,j} C_i\, P_{ij}\,N_j
\label{eq:ym_worldline_amp}
\end{equation}
As in field theory, gauge invariance relies on the algebraic color identities in Eq.~\eqref{eq:color_id_qft}. We would impose precisely the very same generic algebraic identities on kinematic factors $N_j$. Once this could be done, this leads to the proposal
\begin{equation}
A_{\GR}(n) = (-1)^{n}\,\sum_{i,j} N_i\, P_{ij}\,\tilde{N}_j,
\label{eq:gr_worldline_amp}
\end{equation}
as the amplitude in dilaton gravity at $\mathcal{O}(\kappa^{1+2n})$. We replace the polarization in one of the copies as $\tilde{e}_{\nu}$ such that the polarization has a factorized form $e_{\mu\nu} = e_{\mu}\tilde{e}_{\nu}$.

This proposal are motivated by gauge invariance and locality, as in the conventional BCJ double copy. 
Despite other differences, the gauge invariance in YM also relies are the color identities. Therefore we can replace the color factors with kinematic numerators with the same algebraic properties. This guarantees the gauge invariance in YM.
Since the two color copies are on equal footing in bi-adjoint scalar theory, the amplitude in Eq.~\eqref{eq:gr_worldline_amp} has doubly gauge invariance. 
For locality, we find that the pole structure are still given by the usual propagator of radiation field, and by solving the worldline EoM, e.g., Eq.~\eqref{eq:pos_LO} and \eqref{eq:color_LO}.
So keeping the kinematic kernel $P_{ij}$ in bi-adjoint scalar for YM and gravity seems reasonable which we shall verify explicitly.

\subsection{Duality-Satisfying Representations}
To find duality-satisfying representations at next-to-leading order, first consider the color-stripped amplitudes in bi-adjoint scalar theory and YM.
Then we make an local kinematic ansatz $N_i$ for each distinct (dual) color numerators $C_i$ in Table~\ref{table:g5_color}.
Crucially, we demand the ansatze satisfy the same isometries and Jacobi identity as color numerators. Otherwise gauge invariance is not guaranteed later on. For example, the ansatz for $N_1$ should be symmetric under exchange of 2 and 3 from the property of $C_1$;
and the ansatz for $N_4$ should be anti-symmetric under $2\leftrightarrow 3$ as in $C_4$.
A caveat here is that we should only focus on the overall algebraic relation than peeping into the detailed structure.
From the leading-order rules in Ref.~\cite{Goldberger:2016iau}, it is tempting to assume the kinematic ansatz for $C_{4,5,6}$ as $(p_{1,2,3}\cdot e)$ times a universal anti-symmetric function, analogous to $f^{abe} c^a_1 c^b_2 c^e_3$.
However, this demands too much from color-kinematics duality, since the algebraic relations are all we need for gauge invariance.
Thus, we emphasize that only the algebraic relations in Eq.~\eqref{eq:color_id_qft} are imposed.

Matching all color-stripped amplitudes via Eq.~\eqref{eq:bs_worldline_amp} and Eq.~\eqref{eq:ym_worldline_amp} under
\begin{equation}
\tC_i, \, \rightarrow \, N_i,
\end{equation}
then yields a solution with 47 free parameters, corresponding to different generalized gauge degrees of freedom. 
It is not obvious that the solutions exist at all. A particular choice of the duality-satisfying representations is given in Table~\ref{table:bcj_NLO}.
It is straightforward to check that the kinematic counterparts satisfy
\begin{equation}
N_{19}+N_{20}+N_{21}=0 \nonumber
\end{equation}
as the first example of kinematic Jacobi identity beyond field-theory amplitudes!
Note that it only holds under the kinematic constraints, $l_i\cdot p_i =0$, and the on-shell and transverse conditions.

\begin{table}[t]
	\def\arraystretch{1.8}
	\centering
	\begin{tabular}{l l}
		\hline\hline
		$N_{1}\,\,=$ & $(p_1 \cdot p_2)(p_1 \cdot p_3) (p_{1} \cdot e)$ \\
		\hline
		$N_{4}\,\,=$ & 
$\left( (p_1 \cdot \left[p_3, p_2\right] \cdot l_{23})+\frac{1}{2} (\inn{p_2}{p_3}) (\inn{p_1}{q_{23}})\right)
(p_1 \cdot e)+\frac{1}{2}\,(\inn{p_1}{l_{23}})\, (p_1\cdot \left[p_2,p_3 \right] \cdot e)$ \\
\hline
		$N_{7}\,\,=$ & $(p_1 \cdot p_2)(p_2 \cdot p_3) (p_{1} \cdot e)$ \\
		\hline
		$N_{13}=$ & 
		$(p_1 \cdot p_3)\left( (l_{123}\cdot \left[p_1,p_2 \right]\cdot e) -(p_1 \cdot p_2) (\inn{l_2}{e})  \right)$ \\
		\hline
		$N_{19}=$ &
		$\left( (l_1\cdot [p_2,p_3] \cdot l_{23})-\frac{1}{2}(\inn{p_2}{p_3})(\inn{l_{123}}{q_{23}}) \right) (p_1\cdot e)
		+\left(\inn{p_1}{l_{23}}\right) \left({l_{23}} \cdot {\left[p_2,p_3\right]}\cdot e\right)$ \\
		 & 
		$+ \left(p_1 \cdot [p_2,p_3] \cdot l_{23}\right) \left(l_{23}\cdot e\right)
		- (\inn{p_2}{p_3})\left(p_1 \cdot \left[l_2,l_3\right] \cdot e \right)
		-\frac{1}{2}(\inn{l_{23}}{l_{123}})\left(p_1 \cdot \left[p_2,p_3\right] \cdot e \right)
		$
		\\
		\hline\hline
	\end{tabular}
	\caption{An particular example of duality-satisfying representations where $e_{\mu}$ is the YM polarization and $q^{\mu}_{23}= l^\mu_2-l^\mu_3$. 
	We define the tensor $[a,b]^{\mu\nu}= a^{\mu} b^{\nu} - b^{\mu} a^{\nu}$ such that $p\cdot [a,b]\cdot q=(\inn{p}{a})(\inn{b}{q})-(\inn{p}{b})(\inn{a}{q})$. 
	All other numerators are related by permutations. 
}
	\label{table:bcj_NLO}
\end{table}

Using any of the duality-satisfying representations obtained from BS and YM, we are now ready to test the double-copy construction at the next-to-leading order. 
On one hand, we first take the BS amplitude and then substitute the color and dual color factors with
duality-satisfying numerators as instructed by Eq.~\eqref{eq:bcj}.\footnote{The polarization in one of the kinematic copies is replaced with $\tilde{e}_{\mu}$.}
The explicit formulae of the BS amplitude is summarized in Eq.~\eqref{eq:bs_rad_NLO_1}, \eqref{eq:bs_rad_NLO_2}, \eqref{eq:bs_rad_NLO_3} and~\eqref{eq:bs_rad_NLO_4}, and the duality-satisfying numerators are given in Table~\ref{table:bcj_NLO}.
This yields the amplitude in Eq.~\eqref{eq:gr_worldline_amp}.
On the other hand, we calculate the graviton and dilaton amplitudes using the standard EoM in dilaton gravity.
Decomposing the double-copy amplitude via Eq.~\eqref{eq:decomposition}, we arrive at the main result of this paper: \emph{The amplitudes agree precisely between the YM double copy and dilaton gravity.}
It is miraculous that the rather enormous expression in gravity can be summarized into a few equations.

Let us give a few remarks on the duality-satisfying representations we found here. 
Goldberger and Ridgway found amazingly simple rules to match between YM and gravitational radiations~\cite{Goldberger:2016iau}.
The instruction is to replace
color charge with momentum, and the structure constant with three-gluon vertex in YM with appropriate momentum flow. 
Not unexpectedly, it gets a bit more complicated at higher orders.
First, the solutions are far from unique, because numerators by themselves are not physical. The same ambiguity occurs to the BCJ numerators in field theory. 
In fact, we will see that even the leading-order rules are not unique.
At next-to-leading order, the leading-order intuition suggests that $C_4$ in Table~\ref{table:g5_color} should be proportional to $(p_1\cdot e)$ which is in tension with the full range of solutions we find.
Also it is unclear how to interpret a structure constant as a YM three-gluon vertex beyond leading order because the momentum flow is ambiguous.
Therefore we believe the general statement of color-kinematics duality relies on the algebraic relations in Eq.~\eqref{eq:color_id_qft}, 
while the interpretation of color charge as momentum and structure constant as three-gluon vertex in YM are special rules at leading order.

\begin{figure}[t]
	\centering
	\includegraphics[trim={0 15cm 0 0},clip,scale=.4]{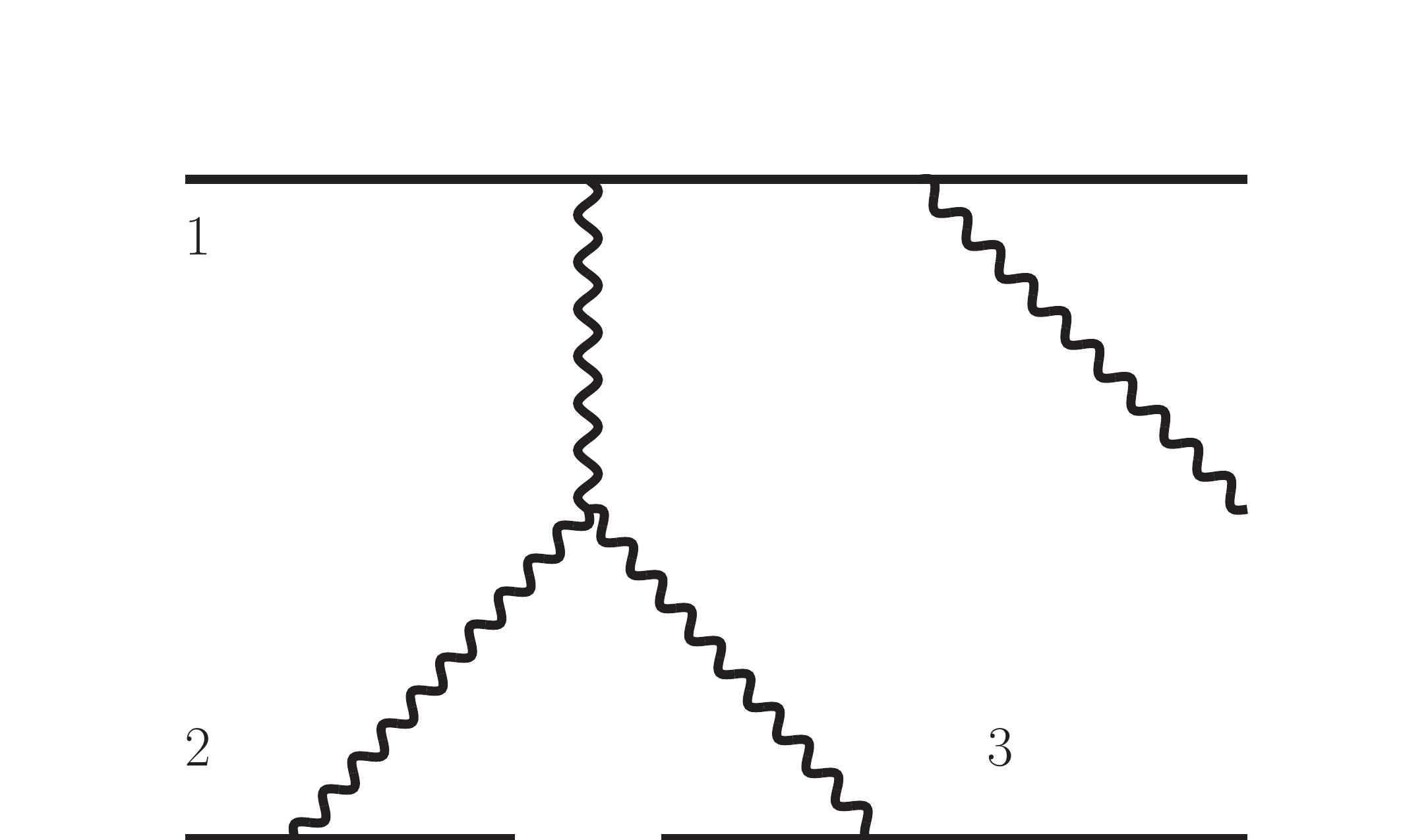}
	\caption{A schematic diagram which is zero is YM but non-zero in (dilaton) gravity. The conventions follow Figure~\ref{fig:self} but the wavy lines are gauge fields in YM, or gravitons in gravity. Particle $1$ absorbs the leading-order radiation from $2$ and $3$ before emitting the final radiation.
		The corresponding color factor is zero by antisymmetry of the structure constant once we identify the color $c_3$ with $c_2$, following the prescription Eq.~\eqref{eq:3_body}. The corresponding diagram is non-zero for gravity.}
	\label{fig:1PN}
\end{figure}

The formulation in Eq.~\eqref{eq:3_body}, treating a two-body system as a three-body system, is very important for double copy to work.
By breaking the degeneracy, we keeps many color factors in YM from vanishing.
For example, the color factor $C_4=(c_1\cdot [c_2,c_3]) c^a_1$, which appears in Figure~\ref{fig:1PN}, vanishes identically once we identify $c_3$ as $c_1$ or $c_2$.
While this treatment seems to complicate YM calculation unnecessarily, this is exactly what we need to make the double copy works, because the very same diagram is not zero in gravity.
Indeed, this is consistent with our rules, where the kinematic counterpart in Table~\ref{table:bcj_NLO} is not zero.
Suppose we consider a two-body system in the beginning, this class of color factors will lead to $C_i P_{ij} N_j=0$ which map to $N_i P_{ij} N_j \neq 0$ under the double copy.
This ``zero-to-nonzero'' phenomena is not special: there are several examples already in multi-loop calculations~\cite{Bern:2012uf}.
So our formulation is actually preferred by keeping those color factors around in the beginning. 
We expect the same behavior to happen at higher orders.

Although the numerators are not unique, it is possible that there is a general pattern for \textit{one particular good choice} of the duality-satisfying representations. Comparing the color and kinematic numerators in Table~\ref{table:g5_color} and~\ref{table:bcj_NLO}, we observe that
\begin{equation}
	c_i\cdot c_j \rightarrow p_i \cdot p_j
\end{equation}
is compatible with color-kinematics duality so far. 
Such pattern, if continuing to hold at higher orders, may lead to a kinematic algebra manifesting the color-kinematics duality, and could pinpoint to the origin of double copy~\cite{Monteiro:2011pc,Monteiro:2013rya,Cheung:2016prv,Cheung:2017yef}.

\subsection{Amplitude Relations}
The color-kinematics duality comes hand-in-hand with two types of relations among physical on-shell amplitudes. The first type is the BCJ amplitude relations among color-stripped amplitudes. The second type is the KLT relations realizing the double copy directly at amplitude level. Since the KLT relations are for physical amplitudes, it allows us to obtain gravity amplitudes without the specification of duality-satisfying representations.
Let us consider them in turn for classical radiation.

We begin by reviewing the BCJ amplitude relations. Consider the color-stripped amplitudes in bi-adjoint scalar theory and YM obtained by projecting the color numerators $C_i$ to a minimal basis.\footnote{We abuse the notation slight by using the same $C_i$ for color factors in a minimal basis}
Using the form in Eq.~\eqref{eq:bs_worldline_amp} and Eq.~\eqref{eq:ym_worldline_amp}, we find
$A_{\BS}(n,C_{i})=P_{ij} \tC_{j}$ and $A_{\YM}(n,C_{i})=P_{ij} N_{j}$. The kernel $P_{ij}$ can be view as a matrix with indices run over minimal basis of color factors. If the kernel is invertible, it would give rise to a unique duality-satisfying representations by matching color-stripped amplitudes in bi-adjoint scalar theory and YM. However, it is often not the case and consistent with the fact that duality-satisfying representations are not unique. This immediately
implies that the kernel $P_{ij}$ has null vectors $v_i$ such that
\begin{equation}
	\sum_{i}\, v_{i}\, A_{\BS,\YM}(n,C_{i}) =0
	\label{eq:bcj_relations}
\end{equation}
as non-trivial relations among color-stripped amplitudes. Note that this relation is universal: it holds for any theories satisfying color-kinematics duality.

For classical radiation, we find one BCJ relation even at leading order
\begin{equation}
\begin{split}
	(\inn{p_1}{l_2}) A(1,C_1)+(\inn{p_2}{l_1}) A(1,C_2)+\frac{1}{2}\left(l^2_1-l^2_2\right)A(1,C_3)=0.
	\label{eq:bcj_LO}
\end{split}
\end{equation}
which holds for both bi-adjoint scalar theory and YM.
A quick way to verify the above relation is from gauge invariance. Consider Eq.~\eqref{eq:gr_worldline_amp} with a longitudinal polarization $\tilde{e}^{\mu} \propto l^\mu_{12}$, the left-hand-side vanishes identically by gauge invariance and $N_i$ reduces to the null vector $v_i$ in above equation.
This relation implies that the leading-order rule~\cite{Goldberger:2016iau}, although being the simplest choice, is not unique. 
We can shift the rules with anything proportional to the above relation without changing the results.

Since double copy relates physical amplitude of various theories, it should be possible to do so directly at the level of amplitudes at least for tree level. Indeed, such construction is the famous KLT relations. Again, consider the amplitudes in double copy form in Eq.~\eqref{eq:ym_worldline_amp} and Eq.~\eqref{eq:gr_worldline_amp}.
Let us ignore that $P_{ij}$ is not invertible for a moment. The gravity amplitude is naively given by
\begin{equation}
\begin{split}
	A_{\GR}(n) &= (-1)^n\,\sum_{i,j} N_k\, P_{ki}\, P^{-1}_{ij}\, P_{jl}\,N_l \\
	&= (-1)^n \,\sum_{i,j} A_{\YM}(n,C_i)\, S_{ij}\, A_{\YM}(n,C_j),
	\label{eq:KLT}
\end{split}
\end{equation}
as a consequence of double copy form with $S_{ij}= P^{-1}_{ij}$.
How to reconcile with the fact that $P_{ij}$, as doubly-stripped amplitudes in bi-adjoint scalar theory, is not invertible in general? 
It turns out the above squaring relation still makes sense by introducing pseudo-inverse~\cite{Boels:2012sy}. Alternatively, we can delete the redundant doubly-stripped amplitudes in bi-adjoint scalar theory due to the BCJ relations. The remaining doubly-stripped amplitudes form a reduced kernel $P_{ij}$, which is invertible by construction. The KLT relations can then be derived straightforwardly.

Concretely, return to the classical radiation at leading order. The doubly-stripped amplitude in Eq.~\eqref{eq:bs_rad_LO} form a $3\times 3$ matrix.
If we choose to delete the third column and row, as instructed by Eq.~\eqref{eq:bcj_LO}, the matrix becomes diagonal and readily invertible. Then the general form in Eq.~\eqref{eq:KLT} reduces to
\begin{equation}
\begin{split}
A_{\GR}(n) &= -\,\sum^{2}_{i,j=1} A_{\YM}(n,C_i) \, S_{ij} \,A_{\YM}(n,C_j)
\label{eq:KLT_LO}
\end{split}
\end{equation}
with $S_{11} = l^2_2 (\inn{p_1}{l_2})^2/(l_2\cdot l_{12}) $, $S_{22} = l^2_1 (\inn{p_2}{l_1})^2/(l_1\cdot l_{12})$, and zero otherwise.
Note that the KLT kernel $S_{ij}$ is not necessarily a local matrix.

Both types of relations generalize at next-to-leading order. First, we find four BCJ relations
\begin{equation}
\begin{split}
0=&(\inn{p_1}{l_{123}}) A(2,C_4)
+(\inn{p_2}{l_{123}}) A(2,C_{5\,})
+(\inn{p_3}{l_{123}}) A(2,C_{6\,})\\
&+(\inn{l_1}{l_{123}}) A(2,C_{19})
+(\inn{l_2}{l_{123}}) A(2,C_{20}) \\
0=&(\inn{p_1}{l_{123}}) A(2,C_8)
+(\inn{p_2}{l_{123}}) A(2,C_{10})
+(\inn{p_3}{l_{123}}) A(2,C_{3}) \\
&-(\inn{l_1}{l_{123}}) A(2,C_{18})
+(\inn{l_2}{l_{123}}) A(2,C_{16})
\end{split}
\label{eq:bcj_NLO}
\end{equation}
and also two more relations from the permutations of the second line. 
This implies there are only 16 independent color-stripped amplitudes at next-to-leading order.
By deleting these redundancies, the kernel $P_{ij}$ is now invertible and yields the KLT relations, which we verify numerically.

By realizing double copy at amplitude level, the KLT relations can be a fast numerical way to generate gravitational radiation at higher orders.
It also enables the construction from generalized double copy~\cite{Bern:2017yxu,Bern:2017ucb,Bern:2018jmv}.
It would be interesting to connect the BCJ and KLT relations to those in the S-matrix with fundamental matter~\cite{Johansson:2014zca,Johansson:2015oia,delaCruz:2015dpa,delaCruz:2016wbr,Naculich:2014naa,Brown:2016hck,Brown:2018wss} which we leave for future work.

\section{Discussions and Outlook}
\label{sec:discussion}
In this paper, we calculate the classical radiation in bi-adjoint scalar theory, Yang-Mills theory, and dilaton gravity to next-to-leading order. 
This is the first gravitational result in the third order of post-Minkowskian expansion.
By comparing BS and YM in the form of Eq.~\eqref{eq:bs_worldline_amp} and Eq.~\eqref{eq:ym_worldline_amp}, we realize color-kinematics duality in Eq.~\eqref{eq:kin_id_qft}.
The resulting double copy amplitude in Eq.~\eqref{eq:gr_worldline_amp} completely agree with explicit calculation in dilaton gravity. 
We also observe the BCJ and KLT relations which are also the signatures of color-kinematics duality.

From S-matrix to classical radiation, the seamless transition of double copy hinges on the gauge invariance and locality---both works surprisingly well in the classical limit.
The gauge invariance in emission amplitudes is based on the on-shell limits of the currents in Eq.~\eqref{eq:master_rad_YM} and Eq.~\eqref{eq:master_gr}. Although the currents look like loop integrands in the S-matrix, they are gauge invariant prior to the integration by our direct calculation. This makes gauge invariance is  preserved under double copy without subtlety. 
Second, locality also holds well by the very same kernel in Eq.~\eqref{eq:bs_worldline_amp} for YM and dilaton gravity. 
Further study along the line of Ref.~\cite{Luna:2017dtq} might make the observations here transparent.
We also find it important to break the degeneracy in color---by considering a generic three-body system---to see the double-copy structure.

Given the extension to broader theories at leading order~\cite{Goldberger:2017ogt,Chester:2017vcz,Li:2018qap,Luna:2017dtq}, it would be interesting to see if the same can be done at next-to-leading order. 
The case with spinning sources~\cite{Goldberger:2017ogt,Li:2018qap} is particularly interesting.
As the calculation becomes more involved, double copy could be even more valuable than in scalar case here.
Second, the anti-symmetric two-form can be appear when spin is non-zero. Confirming double copy including spin and the two-form would complete the full picture.
More drastically, the S-matrix in seemingly different theories, e.g., YM and the non-linear sigma model, are unified under the transmutations in Ref.~\cite{Cheung:2017ems}. Given the success of double copy here, it is possible that the same unifying relations carry through to classical radiation in various theories.

Of course, the most interesting extension would be classical radiation in Einstein gravity. This has been demonstrated at leading order~\cite{Luna:2017dtq}. 
In fact, the modified gauge theory written in Ref.~\cite{Luna:2017dtq} is rather unique by demanding the coupling between the extra scalar and a massive source proportional to mass. It is possible that the same model also works at higher orders. If it succeeds, the results would immediately useful as the first calculation at third post-Minkowskian order in general relativity.

The success at next-to-leading order suggests that the future at higher orders is promising. 
Crucially, worldline formalism trivializes higher order calculations as tree-level-like processes with multiple source insertions, along the line of Eq.~\eqref{eq:3_body}. If we can fully understand such classical processes as tree-level amplitudes in field theory, extending the results in Ref.~\cite{Luna:2017dtq}, then the well-developed techniques in S-matrix will be available. They could teach us how the double copy works at higher order, for broader type of sources, and possibly lead to an all-order proof.

For future applications to LIGO, it relies on how much the double-copy construction can simplify binary black holes especially in post-Newtonian regime. We have demonstrated that the classical limit does not obstruct double copy, at least in the scattering process in post-Minkowskian regime. However, post-Newtonian regime remains quite distinct in many qualitative features~\cite{Goldberger:2004jt}. The time dependence of orbiting binaries would less straightforward as scattering particles.
However, the recent success at leading order \cite{Goldberger:2017vcg} suggests that these complications can be controlled.
Alternatively, the results in post-Minkowskian scheme can be used as for extracting useful information about post-Newtonian systems by using subtraction~\cite{Iwasaki:1971vb,BjerrumBohr:2002kt,Holstein:2008sx,Neill:2013wsa} or effective one-body formalism~\cite{Damour:2016gwp,Bini:2017xzy,Vines:2017hyw,Damour:2017zjx}.
We hope the hidden simplicity found in gravitational S-matrix could eventually shed light on the spacetime ripples from  binary black holes.

\section*{Acknowledgment}
The author would like to thank Clifford Cheung, Walter Goldberger, Henrik Johansson, Aneesh Manohar, Duff Neill, Julio Parra-Martinez, Mikhail Solon, Jan Steinhoff, Jedidiah Thompson, Gabriele Veneziano, and Justin Vines for helpful discussions. He is particularly indebted to Alec Ridgway for many helpful discussions and comments on the manuscript, and to Zvi Bern for helpful discussions, comments on the manuscript, and the continuous encouragement.

\appendix
\section{Notations in Ancillary Files}
\label{app:def}

In all ancillary files, the notations are
\begin{enumerate}
	\item \texttt{p[i]} and \texttt{l[i]} denotes $p_i$ and $l_i$ respectively.
	\item \texttt{nc[i]} and \texttt{nct[i]} is the respective color numerator $C_i$ and $\tC_i$.
	\item \texttt{kk[a,b]} is the Lorentz inner product $a\cdot b$ between two Lorentz vector $a^\mu$ and $b^\mu$, except that we use \texttt{kk[a,b[1]]} for a Lorentz vector $a^{\mu}$.
	\item \texttt{keString[a,eE[1],b]} is a tensor $(a^\mu b^\nu+a^\nu b^\mu)/2$ composed of two Lorentz vector $a^\mu$ and $b^\mu$.
\end{enumerate}

\section{Point Sources at Next-to-leading Order}
\label{app:worldline}
\begin{table}[t]
	\def\arraystretch{1.5}
	\centering
	\begin{tabular}{l l l l}
		\hline\hline
		$C_1=(\inn{c_1}{c_2})(\inn{c_1}{c_3})$ & $C_5=(\inn{c_1}{c_3}) [c_1,c_2]^a$ & $C_9\,\,=[c_1, [c_2,c_3]]^a$ \\
		\hline
		$C_2=(\inn{c_1}{c_2})(\inn{c_2}{c_3})$ & $C_6=(\inn{c_1}{c_2}) [c_3,c_1]^a$ & $C_{10}=[c_2, [c_3,c_1]]^a$ \\ 
		\hline
		$C_3=(\inn{c_1}{c_3})(\inn{c_2}{c_3})$ & $C_7=(\inn{c_2}{c_3}) [c_1,c_2]^a$ & $C_{11}=[c_3, [c_1,c_2]]^a$ \\
		\hline
		$C_4=c_1\cdot [c_2,c_3]$ & 	$C_8=(\inn{c_2}{c_3}) [c_3,c_1]^a$   & \\
		\hline\hline
	\end{tabular}
	\caption{List of the color numerators relevant for worldline degrees of freedom at next-to-leading order. The first column is color singlet and used for the trajectory deviation, while the rest carries an adjoint index and used in the color deviation. The dual color numerators follow similarly. The last three satisfy Jacobi identity $C_9+ C_{10} + C_{11}=0$.}
	\label{table:g4_color}
\end{table}
We give the details of worldline degrees of freedom at next-to-leading order.
First, the color numerators needed for trajectory and color charges are given in Table~\ref{table:g4_color}.
For a point particle labeled by subscript $1$ in bi-adjoint scalar theory, we find the next-to-leading order trajectory deviation is
\begin{equation}
\begin{split}
(p_1\cdot l_{23})^2 \,\delta\mathcal{X}^{\mu}_1(2) =& 
\, C_1 \tC_1 \,
\frac{(l_2\cdot l_3) \left(\left(l_2\cdot p_1\right)^2 l^{\mu }_2 +\left(l_3\cdot p_1\right)^2 l^{\mu }_3 \right)}{l^2_2 l^2_3 \left(l_2\cdot p_1\right)^2 \left(l_3\cdot p_1\right)^2 }
+C_4 \tC_4 \, \frac{2 l_{23}^{\mu}}{l^2_2 l^2_3 l^2_{23} } \\
& -C_2 \tC_2\, \left(\frac{(l_3\cdot l_{23})}{l^2_3 l^2_{23} \left(l_3\cdot p_2\right)^2 }\right)\, l_{23}^{\mu }
-C_3 \tC_3 \, \left(\frac{ (l_2\cdot l_{23})}{l^2_2 l^2_{23} \left(l_2\cdot p_3\right)^2 }\right) l_{23}^{\mu} \\
&+ \left(C_1\tC_4 + C_4\tC_1 \right)\,\frac{ (l_2\cdot p_1) l^{\mu }_2-(l_3\cdot p_1) l^{\mu }_3}{l^2_2 l^2_3 (l_2\cdot p_1) (l_3\cdot p_1) }\\
&-\left(C_2 \tC_4 + C_4 \tC_2\right)\,\frac{l_{23}^{\mu }}{l^2_3 l^2_{23} (l_3\cdot p_2) }
+\left(C_3 \tC_4 + C_4 \tC_3\right)\, \frac{l_{23}^{\mu}}{l^2_2 l^2_{23} (l_2\cdot p_3) } \\
& 
\end{split}
\end{equation}
which is induced by the radiation emitted by worldline 2 and 3, and the color deviation is
\begin{equation}
\begin{split}
(p_1\cdot l_{23})\,\delta\mathcal{C}^{a}_1(2) =& 
-C_5  \frac{(l_2\cdot l_3) \tC_1 + (l_3\cdot p_1) \tC_4}{l^2_2 l^2_3 \left(l_3\cdot p_1\right)^2 }
+C_6\frac{(l_2\cdot l_3) \tC_1 - (l_2\cdot p_1) \tC_4 }{l^2_2 l^2_3 \left(l_2\cdot p_1\right)^2 } \\
&+C_7 \frac{(l_3\cdot l_{23}) \tC_2 + (l_3\cdot p_2)\tC_4 }{l^2_3 l^2_{23} \left(l_3\cdot p_2\right)^2 }
- C_8 \frac{(l_2\cdot l_{23}) \tC_3 - (l_2\cdot p_3)\tC_4 }{l^2_2 l^2_{23} \left(l_2\cdot p_3\right)^2 } \\
&+C_{10} \left(
\frac{\tC_1}{l^2_2 l^2_3 (l_3\cdot p_1)}
-\frac{\tC_2}{l^2_3 l^2_{23} (l_3\cdot p_2) }
+\frac{\tC_3}{l^2_2 l^2_{23} (l_2\cdot p_3) }
+\frac{2\tC_4}{l^2_2 l^2_3 l^2_{23} }
\right) \\
&+C_{11} \left(
-\frac{\tC_1}{l^2_2 l^2_3 (l_2\cdot p_1) }
-\frac{\tC_2}{l^2_3 l^2_{23} (l_3\cdot p_2) }
+\frac{\tC_3}{l^2_2 l^2_{23} (l_2\cdot p_3) }
+\frac{2\tC_4}{l^2_2 l^2_3 l^2_{23} }
\right)
\end{split}
\end{equation}
where we have used $C_9=-C_{10}-C_{11}$ from Jacobi identity 
For YM, we also calculate $\delta\mathcal{X}^{\mu}_1(2)$ and $\delta\mathcal{C}^{a}_1(2)$ which are corrections at next-to-leading order.
The full details are available in the ancillary files \texttt{ym\_pos\_NLO.m} and \texttt{ym\_color\_NLO.m} which involve the same color numerators in Table~\ref{table:g4_color}. The $\delta\mathcal{X}^{\mu}_1(2)$ for a point source in dilaton gravity is also provided in \texttt{gr\_pos\_NLO.m}.

\bibliographystyle{utphys-modified}
\bibliography{classical_BCJ}

\end{document}